\documentclass[useAMS,usenatbib,letterpaper]{mn2e}
\usepackage{graphicx,amsmath,color,amssymb}
\usepackage[pdftitle={The Impact of Galactic Feedback on the CGM},pdfauthor={J. Suresh}]{hyperref}
\usepackage[all]{hypcap}
\usepackage{subfigure}

\newcommand{\adsurl}[1]{\href{#1}{ADS}}

\renewcommand{\thefootnote}{\fnsymbol{footnote}}
   
\newlength{\narrowFigurewidth}
\newlength{\Figurewidth}
\newlength{\newFigurewidth}
\newlength{\wideFigurewidth}
\newlength{\widerFigurewidth}
\newcommand{\eq}[1]
  {Eq.~(\ref{eqn:#1})}
\newcommand{\sect}[1]
  {Section~\ref{sec:#1}}
\newcommand{\tabl}[1]
  {{\mbox Table~\ref{table:#1}}}
\newcommand{\fig}[1]
  {Fig.~\ref{fig:#1}}

\newcommand{\Msun}{\, M_\odot}
\newcommand{\Msunh}{\, h^{-1} M_\odot}

\newcommand{\ion}[1]{{\small #1\,}}

\title[The Impact of Galactic Feedback on the Circumgalactic Medium]{The Impact of Galactic Feedback on the Circumgalactic Medium}

\author[Joshua Suresh et al.]
{\parbox{\textwidth}{Joshua Suresh,$^{1}$\thanks{E-mail: \texttt{jsuresh@cfa.harvard.edu}}
Simeon Bird,$^{2}$
Mark Vogelsberger,$^{3}$
Shy Genel,$^{1}$
Paul Torrey,$^{3,4}$
Debora Sijacki,$^{5}$
Volker Springel$^{6,7}$ and
Lars Hernquist$^{1}$}\vspace{0.4cm}\\
\parbox{\textwidth}{$^1$Harvard-Smithsonian Center for Astrophysics, 60 Garden Street, Cambridge, MA, 02138, USA\\
$^{2}$Carnegie Mellon University, 5000 Forbes Avenue, Pittsburgh, PA 15213, USA \\
$^{3}$Department of Physics, Kavli Institute for Astrophysics and Space Research, Massachusetts Institute of Technology,
Cambridge, MA 02139, USA\\
$^{4}$Caltech, TAPIR, Mailcode 350-17, California Institute of Technology, Pasadena, CA 91125, USA\\
$^{5}$Institute of Astronomy and Kavli Institute for Cosmology, University of Cambridge, Madingley Road, Cambridge CB3 0HA, UK\\
$^{6}$Heidelberg Institute for Theoretical Studies, Schloss-Wolfsbrunnenweg 35, 69118 Heidelberg, Germany\\
$^{7}$Zentrum f\"{u}r Astronomie der Universit\"{a}t Heidelberg, ARI, M\"{o}nchhofstr. 12-14, 69120 Heidelberg, Germany}}

\begin{document}

\pagenumbering{alph}
\date{}

\maketitle
\pagerange{\pageref{firstpage}--\pageref{lastpage}} \pubyear{2012}

\pagenumbering{arabic}
\label{firstpage}

\begin{abstract}
Galactic feedback strongly affects the way galactic environments are enriched.  We examine this connection by performing a suite of cosmological hydrodynamic simulations, exploring a range of parameters based on the galaxy formation model developed in \cite{2013MNRAS.436.3031V} (henceforth V13).  We examine the effects of AGN feedback, wind mass loading, wind specific energy, and wind metal-loading on the properties of the circumgalactic medium (CGM) of galaxies with $M_\text{halo} > 10^{11} \Msun$. Note that while the V13 model was tuned to match observations including the stellar mass function, no explicit tuning was done for the CGM. The wind energy per unit outflow mass has the most significant effect on the CGM enrichment.  High energy winds launch metals far beyond the virial radius. AGN feedback also has a significant effect, but only at $z < 3$.  We compare to high redshift H\ion{I} and C\ion{IV} observations. All our simulations produce the observed number of Damped Lyman-$\alpha$ Absorbers. At lower column density, several of our simulations produce enough Lyman Limit Systems (LLS) $100$ kpc from the galaxy, but in all cases the LLS abundance falls off with distance faster than observations, with too few LLS at $200$ kpc. Further, in all models the C\ion{IV} abundance drops off too sharply with distance, with too little C\ion{IV} $100$-$200$ kpc from the galaxy.  Higher energy wind models produce more extended C\ion{IV} but also produce less stars, in tension with star-formation rate density observations.  This highlights the fact that circumgalactic observations are a strong constraint on galactic feedback models.
\end{abstract}


\section{Introduction}
\renewcommand{\thefootnote}{\fnsymbol{footnote}}

It is well-established that without efficient feedback, simulated galaxies are too massive and form too many stars.  A number of authors \cite[e.g.,][]{2010MNRAS.402.1536S,2011MNRAS.415...11D,2013MNRAS.428.2966P,2013MNRAS.436.3031V} have found that by tuning feedback in hydrodynamic simulations including strong supernova feedback and AGN, good agreement can be achieved with observed stellar galactic properties such as the galaxy stellar mass function, the cosmic star formation rate density, and the Tully-Fisher relation.
However, despite the successes of feedback implementations in recent galaxy formation models, feedback itself remains poorly understood.  There is uncertainty about how best to implement galactic feedback into simulations, since the relevant baryonic processes, such as star formation or black hole accretion, are not directly resolvable in cosmological galaxy formation simulations \citep[although currently it is possible to resolve GMC scales and smaller in simulations of individual halos, e.g.][]{Hopkins:2013ua}. Given this uncertainty, a number of competing models have arisen which are able to roughly reproduce several properties of the stellar component of galaxies. One way of breaking this degeneracy 
and distinguishing between the competing models is to consider additional observables sensitive to different properties of the feedback implementations, such 
as the distribution of gas in the circumgalactic medium (CGM). 

The basic picture of the CGM has evolved over time.  
Early theoretical work argued that the properties of the CGM were determined largely by the competition between cooling and accretion, and whether or not a virial shock can form \citep{1991ApJ...379...52W,1978MNRAS.183..341W,1977MNRAS.179..541R,1977ApJ...211..638S}.  
More recent studies have revealed a complex three dimensional structure, intimately entwining gas from accretion and outflows, and depending on the properties of both \citep[e.g.,][]{2012ApJ...760...50S}.
Observations suggest that the CGM is enriched by metals above the background level of the IGM on a rough scale of a few hundred kpc around a galaxy \citep[e.g.,][]{2006ApJ...637..648S,2010ApJ...717..289S}.  For example, \cite{2011Sci...334..948T} showed that many galaxies at late times are surrounded by large halos of hot, enriched gas, visible in O\ion{VI} absorption lines.  Simulations in turn connect this enrichment to feedback-driven outflows 
\citep[e.g.,][]{2006MNRAS.373.1265O,2010MNRAS.409..132W,2012MNRAS.420..829O,2013MNRAS.430.1548H,2013MNRAS.tmp..798B}.

The theoretical picture of the enrichment of the CGM in the literature is somewhat uncertain.  For example, \cite{2012ApJ...760...50S,2013ApJ...765...89S} find that metal-enriched gas essentially free-flows out of the halo, with much of the IGM being enriched by outflows from small halos as early as $z > 5$ \citep[see also][]{2014ApJ...792...99S}.  Conversely, \cite{Ford:2013wk}, find a strong circumgalactic fountain at low redshift, with virtually all CGM metals having been in an outflow at some point, and the majority having been in a recent outflow.    Furthermore, whereas the warm-hot metals in the \cite{2012ApJ...760...50S} model were mostly recently ejected in a hot outflow, the warm-hot metals in \cite{Ford:2013wk} tend to be much older (age $> 1 $ Gyr).  This implies that the differences in wind temperatures and feedback implementation may change the interpretation of observed CGM absorption lines.  For example, \cite{2012ApJ...760...50S,2013ApJ...765...89S} found that the lowest overdensities are enriched by relatively cold gas, whereas Ford et al. find exactly the opposite trend; the lower the overdensity the higher the ionization state of the gas \citep[][]{Ford:2013wk,2013MNRAS.432...89F}.  By comparing different models with different wind properties, we will show that many of these discrepancies in CGM properties may, in fact, stem from the different choices of galactic feedback implementation.

In this paper, we will focus on the impact of galactic feedback on the circumgalactic medium.  We begin with a comprehensive galaxy formation model explained in detail in \cite{2013MNRAS.436.3031V} (henceforth V13), which is shown in V13 and \cite{2014MNRAS.438.1985T} to achieve good agreement with a variety of galactic stellar, ISM, and black hole properties, but was not tuned for the CGM. We consider the impact of AGN feedback, galactic wind specific energy (energy per unit wind mass), mass loading, thermal/kinetic energy partition, and metal loading.  Using the same cosmological initial conditions for each physics model, we compare the circumgalactic gas and metal properties around a statistical sample of galaxies at high redshift ($z \geq 2$), near the peak of cosmic star formation and AGN growth.

Besides an analysis of a small number of zoom-simulations of Milky-Way sized galaxies \citep{2014MNRAS.442.3745M}, this is the first systematic study of the CGM carried out with a moving-mesh hydrodynamics code, \textsc{arepo}  \citep{Springel:2010hx}.  \textsc{arepo} combines the strengths of both smoothed particle hydrodynamics and Eulerian grid codes, and has unique advantages as a tool for studying the CGM.
In particular, it allows metals to passively advect between fluid elements, while the ability of the mesh cells to move with the bulk motion of the fluid minimises overmixing \citep{Springel:2010hx,2013MNRAS.435.1426G}. It accurately resolves fluid instabilities
\citep[][]{2012MNRAS.424.2999S,2012MNRAS.427.2224T,2014MNRAS.442.1992H}, which can significantly alter the properties of 
the CGM \citep[][]{2012MNRAS.427.2224T,2013MNRAS.429.3353N}.

\begin{table*}
\begin{tabular}{l|l|c}
\hline
Name & Wind Properties & AGN Feedback?  \\
\hline
\hline
Fiducial & Fiducial & Yes \\
No AGN & Fiducial & No \\
High Mass-Loading & Mass-loading doubled & Yes \\
Faster Winds & Wind speed increased by factor of $\sqrt{2}$ \textsuperscript{*}  & Yes \\
Fixed-E Hot Winds & Heated winds with same specific wind energy as fiducial \textsuperscript{**}& Yes \\
Fixed-v Hot Winds & Heated winds with same wind speed as fiducial.\textsuperscript{***}   & Yes \\
Pristine Winds & Winds launched with zero metallicity & Yes \\
Fully Enriched Winds & Winds launched with same metallicity as wind-launching region & Yes \\
\hline
\multicolumn{3}{l}{\textsuperscript{*}\footnotesize{The total specific wind energy is twice that of the Fiducial run, and equal to that of the Fixed-v Hot Winds run.}}\\
\multicolumn{3}{l}{\textsuperscript{**}\footnotesize{Since the specific wind energy is fixed and the thermal energy is set to be half of the total wind energy, these winds}}\\
\multicolumn{3}{l}{\hspace{4mm}\footnotesize{are 30\% slower than the Fiducial run.}}\\
\multicolumn{3}{l}{\textsuperscript{***}\footnotesize{The thermal energy is set to be half of the total wind energy, but with the same kinetic energy as the Fiducial winds.}}\\
\multicolumn{3}{l}{\hspace{4mm}\footnotesize{Hence the total specific wind energy is twice that of the Fiducial run.  Since the specific wind energy is doubled,}}\\
\multicolumn{3}{l}{\hspace{4mm}\footnotesize{energy conservation mandates that the mass loading is halved.}}\\
\end{tabular}
\caption{Simulation Properties}
\label{table:sims}
\end{table*}

This paper is structured as follows. In \sect{methods}, we discuss the numerical methods used in this study, including the varying physics implementations used in our different feedback models.  In \sect{results} we compare the properties of the CGM around a statistical sample of galaxies in a cosmological volume at $z=2$ among the different feedback models.  We continue in \sect{obs} by comparing the models with observations of H\ion{I} covering fractions and C\ion{IV} column densities.  Finally, we conclude with a summary and discussion of our findings in \sect{discussion}.

\section{Methods}
\label{sec:methods}

In this study, we use the moving-mesh code \textsc{arepo} \citep{Springel:2010hx}.  \textsc{arepo} employs a finite-volume scheme over a grid of cells defined using a Voronoi tesselation of space.
The mesh-generating points which generate the tesselation move approximately with the bulk motion of the fluid flow, and are sized so that each cell contains a roughly equal amount of mass.
 \citep{Springel:2010hx,2012MNRAS.425.3024V,2012MNRAS.424.2999S,2012MNRAS.425.2027K,2012MNRAS.427.2224T}.  
 
All simulations used in this paper have identical initial conditions, which assume a WMAP-7 cosmology 
($\Omega_{\Lambda,0} = 0.73$, $\Omega_{m,0} = 0.27$, $\Omega_{b,0} = 0.0456$, $\sigma_8 = 0.81$ and $h = 0.704$).  We use a periodic box of side length 25 $h^{-1}$ Mpc, with $256^3$ gas cells (and an equal number of dark matter particles).  The dark matter gravitational softening length is set to 2 $h^{-1}$ comoving kpc (ckpc), while stars and black holes have a softening length of 1 $h^{-1}$ ckpc.  Gas cells use an adaptive smoothing length which is tied to the radius of the cell, with a minimum smoothing length of 1 $h^{-1}$ ckpc.  We use a (de-)refinement scheme which keeps gas cell masses close to a fixed target gas cell mass of $1.25\times 10^7\Msunh$ \cite[for more details on this  scheme see][]{2012MNRAS.425.3024V}.  The dark matter particle mass is fixed at $5.86\times 10^7\Msunh$.

\subsection{Fiducial Galaxy Formation Model}
\label{sec:GFM}
For details of the numerical and feedback parameters used in this paper, we refer the reader to V13.  Here we briefly review some salient features.  
Primordial and metal line cooling are included in the presence of a time-varying UV background \citep{Katz:1996tu,2009MNRAS.399..574W,2009ApJ...703.1416F}, including gas self-shielding following \cite{2013MNRAS.430.2427R}.  Star particles are formed using a two-phase ISM model \citep{2003MNRAS.339..289S}, and we include a model for stellar evolution and chemical enrichment described in V13, which explicitly tracks 9 elements: H, He, C, N, O, Ne, Mg, Si, Fe.
Winds are launched from star-forming regions, with the wind speed fiducially set to be 3.7 times the 1D local halo velocity dispersion.  In order to ensure that the winds escape the dense ISM, wind particles are hydrodynamically decoupled and allowed to propagate until they either travel for a set time, or reach sufficiently low-density regions.  The wind particle's mass, momentum, energy, and metals are then deposited into the nearest gas cell.  Each wind particle is assigned a metallicity equal to some fraction of the metallicity of the ambient ISM from which it was launched.  The fiducial wind metal-loading factor of 0.4 was chosen to allow the simulations to match the galaxy mass-metallicity relation.

In the runs which include AGN feedback, black hole (BH) particles are seeded at early times and allowed to grow through accretion and merging events.  Since the scale of the horizon is always well below our resolution, we implement sub-grid feedback models based primarily on the methods developed in earlier studies \citep{2005Natur.433..604D,2008ApJ...676...33D,2005MNRAS.361..776S,2007MNRAS.380..877S}.  BH feedback has two modes, depending on the accretion rate.  At high accretion rates, the BH enters ``quasar-mode", where a fraction of the accretion energy is thermally coupled to the nearby gas.  At lower accretion rates, the BH enters ``radio-mode" feedback, which mimics observations of BH radio jets injecting energy at larger scales by inflating hot bubbles which are hydrodynamically buoyant.  In our implementation, this is carried out by injecting energy in randomly-distributed bubbles at a distance offset from the center of the galaxy.  Finally, we approximate the effect of AGN irradiation by modifying the heating and cooling rates for gas cells close to the BH.

An important aspect of this work is the study of metals and the metal-enrichment of the CGM due to galactic feedback processes. Each star particle which forms in our simulation corresponds to a single-age stellar population (SSP), which generates metals over time depending on its IMF and age (for example, a SSP with a top-heavy IMF will deposit its metals at a higher rate as the more massive stars evolve faster).  We use a Chabrier IMF in all runs.  Enrichment from AGB stellar mass loss, core-collapse supernovae events, and Type Ia supernovae are included, as these are the primary channels by which a stellar population's metal production enriches the nearby ISM (see V13 for more information on assumed yields from each channel).  As metals are generated over time by the star particle, they are deposited into nearby gas cells, after which the metals are treated as a passive tracer which is hydrodynamically advected with the gas.  Feedback processes thus transport metals into the CGM by ejecting enriched gas from the galaxy into the halo.

\subsection{Feedback Variations}

Our results are based on eight different simulations, summarized in \tabl{sims}.  The ``Fiducial'' simulation is the full physics model used in V13 and the much larger ``Illustris'' simulation \citep{2014Natur.509..177V,2014arXiv1405.2921V,2014arXiv1405.3749G}, which includes both galactic winds and AGN feedback.  The ``No AGN'' simulation retains the same galactic wind scheme but does not include AGN feedback.  All other runs keep the same AGN feedback model as used in the Fiducial run (see \sect{GFM} for more details), and consider variations of the galactic wind implementation which we discuss below.  

We investigate two wind-heating prescriptions: the first is the ``Fixed-E Hot Winds'' model, which keeps the galactic wind specific energy (energy per unit wind mass) fixed while injecting 50\% of the supernova energy into heating the winds, \citep[as in][]{2014MNRAS.437.1750M}.  This results in a slower wind, since the total wind energy (and mass-loading) is fixed.  The second wind-heating prescription is the ``Fixed-v Hot Winds'' model, which keeps the wind speed fixed, while again injecting 50\% of the supernova energy into heating the winds.  Since in this case the specific wind energy is higher, energy conservation mandates that the mass-loading in these winds is lower than the Fiducial run.  For more information on our wind heating implementation, and how hot the winds are in our models, please see the Appendix.

The ``Faster Winds'' model has a higher wind speed by a factor of $\sqrt{2}$ (which gives a specific wind energy that is twice as high as the Fiducial model).  The specific wind energy is equal to that of the ``Fixed-v Hot Winds'', except in the latter case the extra wind energy is in the thermal component.  Note that this means there are two pairs of models (Fiducial and Fixed-E Hot Winds; Fixed-v Hot Winds and Faster Winds) which have the same specific wind energy, while varying the partition between kinetic and thermal energy.  This allows us to examine how the wind energy partition affects the CGM, at two different wind-energy scales.  

The ``High Mass-Loading'' model doubles the mass-loading of the galactic winds, without changing their specific energy (thereby doubling the total energy per unit SFR that goes into the wind).  Note that it is the same as the ``stronger winds'' model in V13. 

We include two models which vary the metal-loading of winds.  The Fiducial model has a metal-loading factor in the winds of 0.4, chosen to match the galaxy mass-metallicity relation (see V13).  Algorithmically, this means that the ejected wind elements are launched with a metal content which is 40\% of the metallicity of the wind-launching ISM region.  We investigate the importance of this choice by including two models which have zero-metallicity "pristine" winds, and fully-enriched (metal-loading of 100\%) winds.  

We do not include a ``no-feedback'' model (with no winds or AGN) since, as seen in V13, neglecting efficient feedback results in a huge overproduction of stars and metals, which make the resulting CGM properties incorrect and difficult to interpret.   

\subsection{Identifying Galaxies}
Dark matter halos are identified using the \textsc{subfind} algorithm \citep{2001MNRAS.328..726S}, with a friends-of-friends (FoF) group-finding algorithm, using a linking length of 0.2 times the mean particle separation.  In this paper we limit our analysis to ``central galaxies", or the galaxies which inhabit the largest halo of each FoF group.  This allows us to incorporate the effects of any satellite galaxies on the shared CGM (e.g. small satellites helping to enrich the CGM through stellar outflows or tidal stripping).  We look at all central halos of mass  $> 10^{11} \Msun$, which corresponds to a minimally resolved halo of about 1500 gas cells.  We define the virial radius of each central galaxy as the radius within which the average density exceeds 200 times the critical density (we will refer to this radius as $R_{200}$).

\subsection{Computing column densities and covering fractions}
\label{sec:fc}

In \sect{obs} we compute column densities of Lyman Limit Systems (LLS) and Damped Lyman-$\alpha$ Absorbers (DLAs) around galaxies, as well as column densities of C\ion{IV}.  First we compute the total mass of the desired ion (H\ion{I} or C\ion{IV}) in each gas cell.  The neutral fraction of hydrogen is computed inside \textsc{arepo} using the cooling network \citep{Katz:1996tu} in conjunction with the self-shielding prescription given by \cite{2013MNRAS.430.2427R}.  For other ions, such as C\ion{IV}, we use Cloudy version 13.02 \citep{2013RMxAA..49..137F} to find the ionisation state of each gas resolution element as a function of temperature and density. Cloudy is run in single-zone mode, which assumes that the density and temperature is constant within each element. Photoionisation is included with the same background radiation as in the rest of the simulation. Our method for computing ionisation states is described further in \cite{2014arXiv1407.7858B}, and our implementation is publically available at \url{https://github.com/sbird/cloudy_tables}.

The ionic content of each gas element within 300 proper kpc in front of and behind the galaxy is then projected onto a uniform grid covering 300 proper kpc around each halo (see \fig{grids} for an example). The size of the grid cells scales with redshift, since the optimal grid cell size is comparable to the size of the smallest gas elements, and the cells have a smaller proper size at later times.  Roughly speaking, each grid cell is about $1$ kpc across.  Interpolation is performed using an SPH kernel with size corresponding to the gas cell size as in \cite{2013MNRAS.429.3341B}.  We take each grid cell to be a single sight line, analogous to observations of quasar sightlines at random locations around (uncorrelated) foreground galaxies \cite{1996ApJ...457L..57K}.  The column density is the total mass of the ionic species within the grid cell, divided by the cross-sectional area of the pixel. 

The covering fraction is defined to be the fraction of all sight lines in a desired radial range which have a column density in the range of interest ($10^{17} < N_\text{HI} < 10^{20.3}$ for LLS and $N_\text{HI} > 10^{20.3}$ for DLAs).  These covering fractions do depend on the direction of the sight-lines taken.  However, we have confirmed that we have a sufficiently large sample of galaxies that the distribution of covering fractions is unaffected by taking different sight line directions.

\begin{figure}
\includegraphics[width=\newFigurewidth]{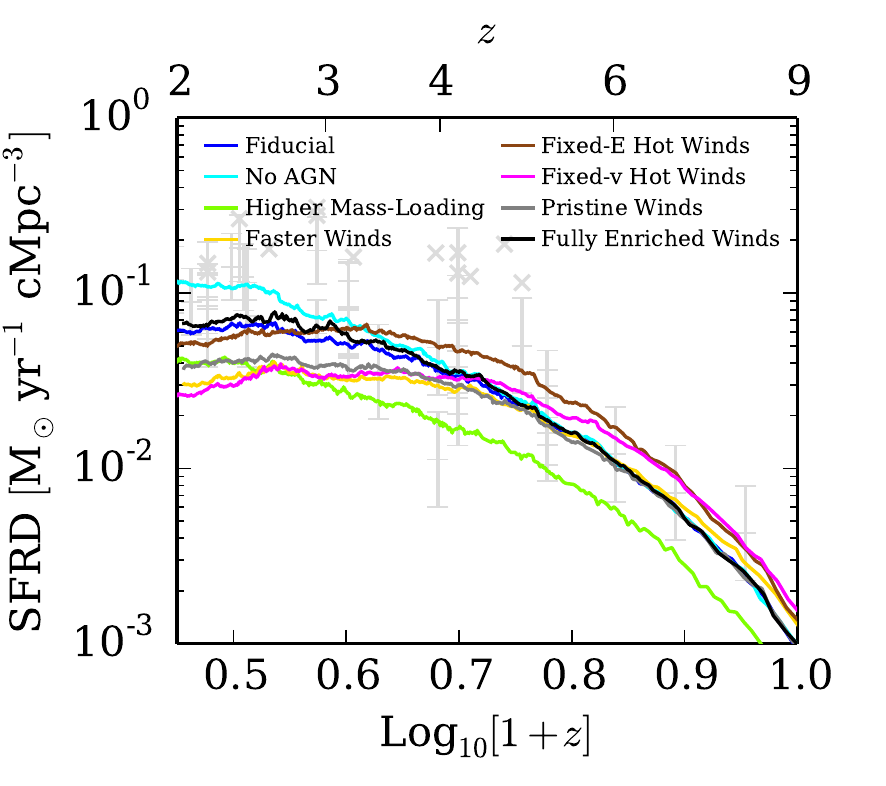}
\label{fig:SFR}
\caption{Cosmic star formation rate density (SFRD) within the full box, as a function of redshift.  Observational points shown in background are from \protect\cite{2006ApJ...651..142H} and \protect\cite{2013ApJ...770...57B}.  AGN feedback does not have a noticeable effect until $z \lesssim 3.5$, after which it significantly reduces cosmic star formation.  High wind mass-loading inhibits star formation at all redshifts (though the ejected gas can fountain back at late times; see V13).  Winds with higher specific energy (in this case, the Faster Winds and Fixed-v Hot Winds) have a correspondingly lower mass loading, which makes them less efficient at slowing star formation at early times.  However, the gas ejected by these higher energy winds is more likely to escape entirely, limiting fountain-triggered star formation at later times.  At fixed specific wind energy, the thermal energy component is less effective at stopping star formation than the kinetic energy component, since the thermal energy can be radiated away promptly.  Wind metal-loading also has an effect on the cosmic SFRD by moderating how effective metal line cooling in the halo is, which modifies the strength of the galactic fountain at later times.}
\end{figure}

\begin{figure*}
\includegraphics[width=\widerFigurewidth]{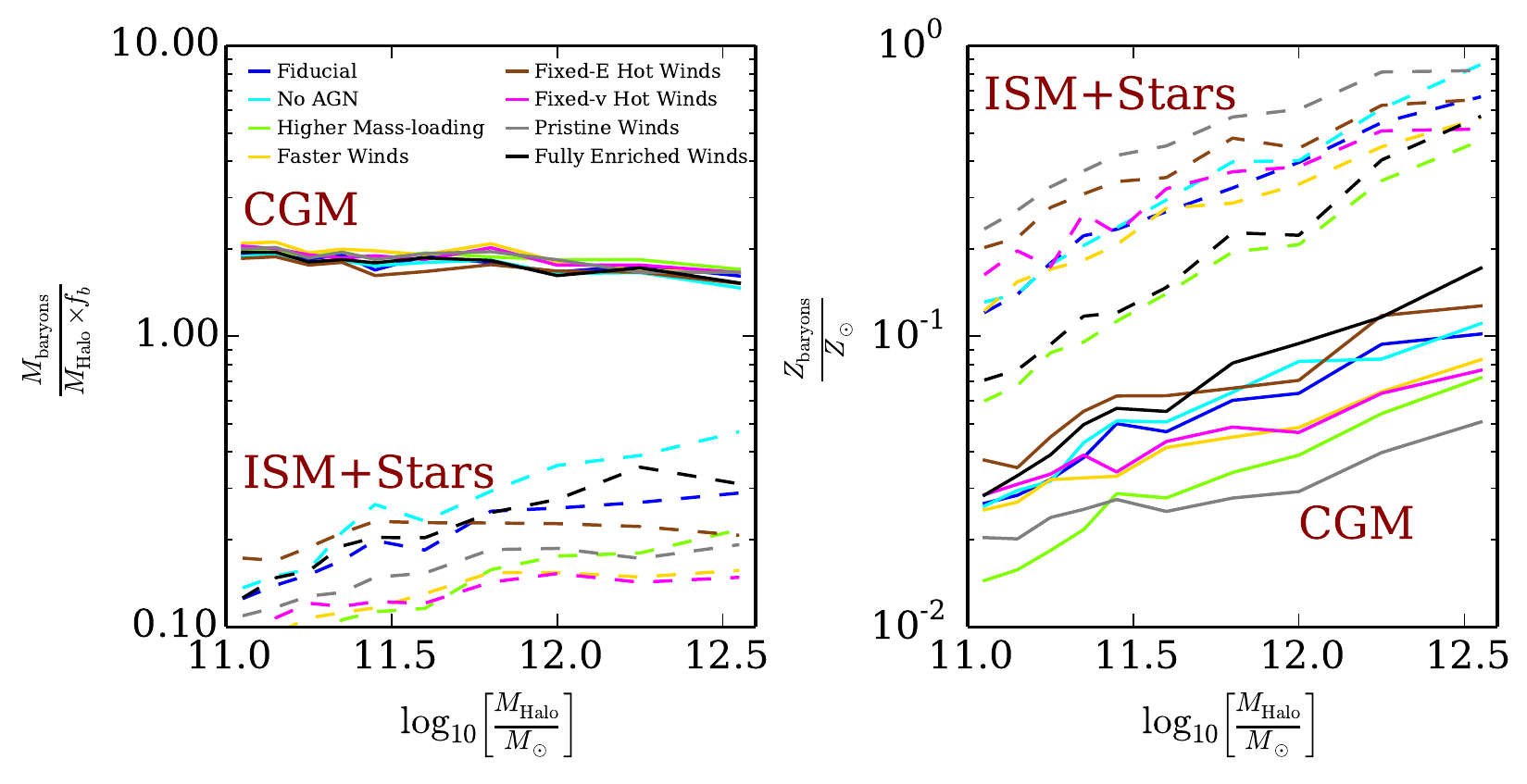}
\label{fig:massmet_budget}
\caption{The total mass (left panel) and metal content (right panel) in the CGM and galaxies at $z=2$.  Curves show the median for all halos in the given mass bin; solid lines show the CGM gas, while dashed lines show the galactic component (both stars and ISM gas).  The left panel shows the mass, in units of the total expected baryon content of the halo, $f_b M_\text{halo} = \frac{\Omega_b}{\Omega_m} M_\text{halo}$.  The mass of the CGM is essentially unaffected by variations in the galactic feedback, while the mass in the galactic component varies significantly (see also \fig{SFR}).  The right panel shows the metal content, in units of Solar metallicity.  Wind-heating has little effect on the total metallicity of the CGM or galaxy.  The high mass-loading has lower galactic and CGM metallicity due to lower overall metal production from suppressed star formation.  AGN have little effect on the total metal content of the CGM.  Higher wind metal-loading acts to drain more metals out of the galaxy into the CGM; we find that fully metal-loaded winds result in a galaxy metallicity which is $2-4$ times lower than pristine winds.
}
\end{figure*}

\section{Simulation Results}
\label{sec:results}

We have checked that our CGM results are reasonably numerically converged by running the code with the same initial conditions at half and double the spatial resolution for the Fiducial Run (8 times fewer/more particles, respectively).  See \sect{converge} for details.  V13 and \cite{2014MNRAS.438.1985T} show the corresponding convergence properties for the galaxies.

\subsection{Star Formation}
\label{sec:sf}
Since metal production is directly tied to star formation, it is important to understand how the different feedback schemes affect star formation rates within each model.  For example, it is possible that a lower CGM metallicity in a given model is driven by a lower metal-production rate, originating from a low star formation rate---indeed this is the case in the High Mass-loading simulation, as we will see below.  For a more detailed analysis of the impact of feedback on star formation, in the context of the current model and code, we refer the reader to V13.

\fig{SFR} shows the total star formation rate density within the full simulation volume, as a function of redshift.  AGN feedback does not have a noticeable effect until $z \lesssim 3.5$, after which it significantly reduces cosmic star formation.  The run with high mass-loading inhibits star formation at high redshifts (though as shown in V13, the ejected gas does not completely escape, and there is a strong galactic fountain at $z < 2$).  

Both Fixed-v Hot Winds and Faster Winds have a higher specific wind energy than the Fiducial Run (Fixed-v Hot Winds puts the extra energy into thermal energy, while Faster Winds puts it into kinetic energy).  As such, both of these runs have a correspondingly lower mass loading, which makes them less efficient at slowing star formation at early times.  However, the smaller amount of gas ejected by these higher-energy winds is more likely to escape the halo entirely, which reduces fountain-triggered star formation at later times (after about $z \sim 4$).

The Fixed-E Hot Winds, which has the same specific wind energy as the Fiducial run, shows that even at a fixed specific wind energy, the thermal component of the energy is less effective at stopping star formation.  This is due to the fact that thermal energy may be radiated away, whereas kinetic energy cannot be directly radiated away.  However, galactic winds in these simulations often shock and thermalize once they reach the halo gas, after which the kinetic energy in the winds is partly converted to thermal energy and can be subsequently radiated away.  Wind thermalization is discussed further in later sections.

The Pristine Winds and Fully Enriched Winds runs show that even at fixed wind speed and energy, the metal-loading of the winds has an effect on the star formation at $z \lesssim 4$.  Winds with high metal-loading greatly enrich the bound halo gas (see \fig{radprof}), which can then cool more efficiently through metal line cooling.  This enhanced cooling boosts the rate of gas condensation back onto the central galaxy, eventually enhancing star formation.  Note that we have confirmed the importance of metal line cooling in the CGM by comparing to a run with only primordial cooling.

\subsection{Mass and Metal Content of the CGM}
\label{sec:mass_met}
In this section we consider the CGM to be bounded by a sphere 3 virial radii ($R_{200}$) from the galactic centre.  The virial radius is used as a simple way to scale the size of the included region with the halo mass, as opposed to taking a fixed radius around both small and massive halos.  We subdivide the gas by defining the ISM as gas with a density above the star formation threshold ($\rho_\text{SF}$ = 0.13 cm$^{-3}$), and the CGM to be all non-starforming gas within 3 $R_{200}$.  In \fig{massmet_budget} we plot the total gas mass and metallicity of galaxies and their respective CGM, as a function of halo mass.  Note that in this figure, the galactic component includes both the central and satellite stars and ISM (though the central galaxy dominates).

\begin{figure}
\includegraphics[width=\Figurewidth]{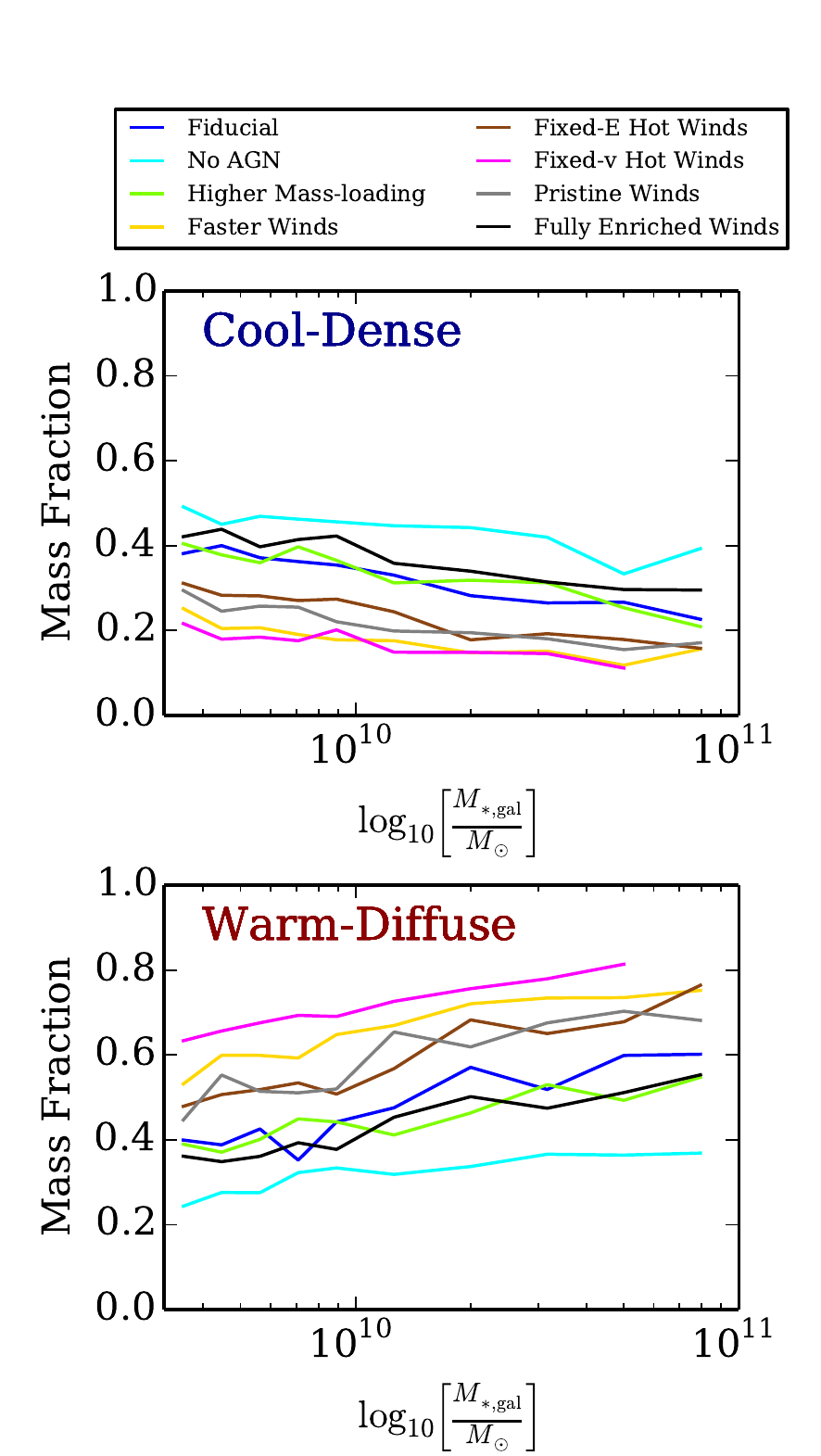}
\label{fig:mass}
\caption{The mass fractions of the CGM phases at $z=2$. The cool-dense phase, though comprising very little of the CGM volume, is about 15-50\% of the mass.  The mass fractions in the CGM vary much less with feedback than the metal fractions of the CGM (compare with \fig{met}). We omit the cool-diffuse and warm-dense phases, which are essentially fixed at about 10\% across all models.}
\end{figure}
\begin{figure}
\includegraphics[width=\Figurewidth]{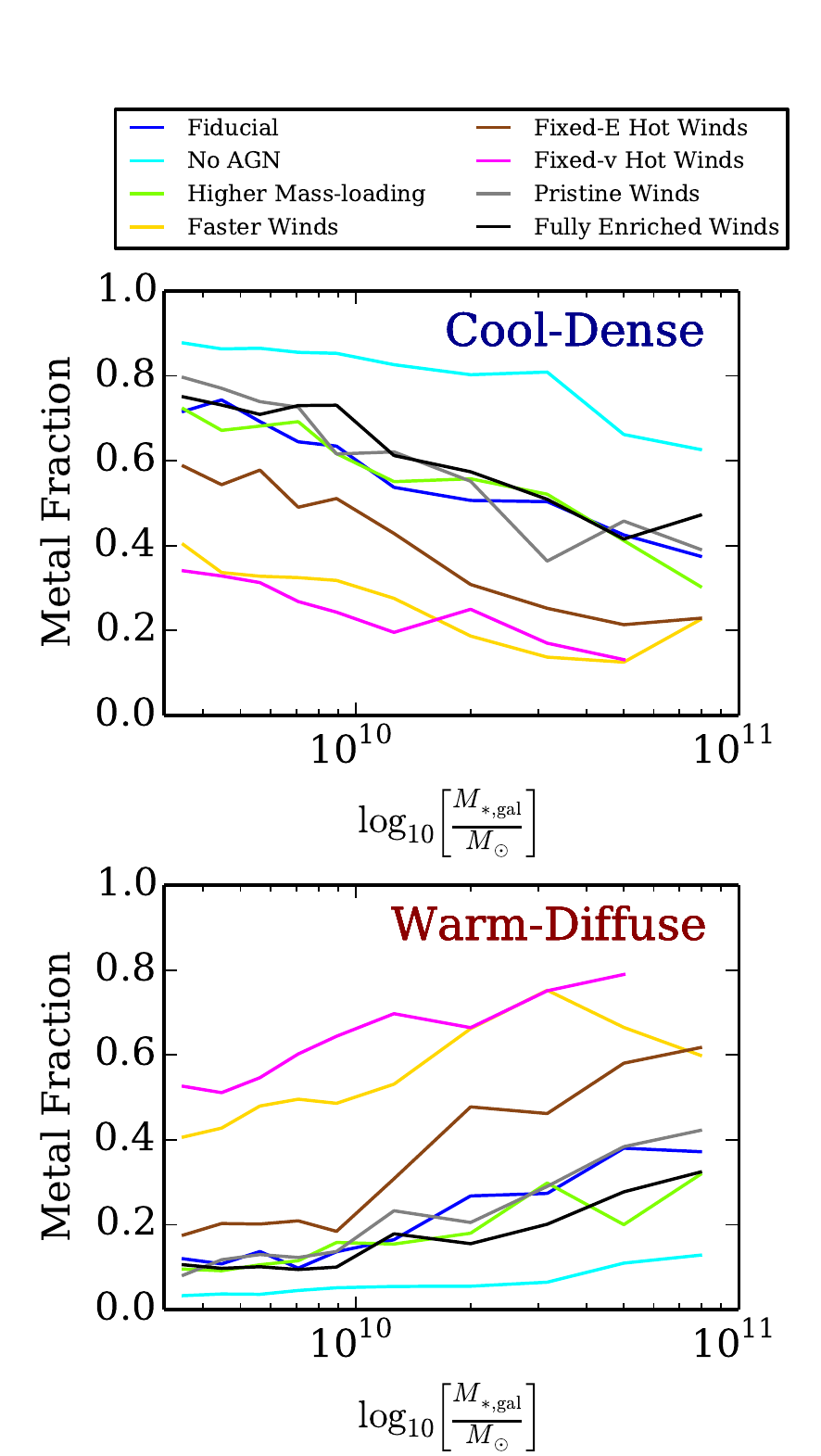}
\label{fig:met}
\caption{The metal fractions of the CGM phases at $z=2$.  The metal budgets are significantly different with feedback, compared with the mass budgets (compare with \fig{mass}).  AGN effectively heat the CGM, transferring metals from the cool-dense phase into the warm-diffuse phase.  The thermal/kinetic energy partition of the winds has some effect on the CGM metal budget (compare Fiducial and Fixed-E hot winds).  However, the the total specific wind energy has significantly stronger effect, with higher energy winds (Faster Winds and Fixed-v Hot Winds models) heating the CGM metals out of the cool-dense phase into the warm-diffuse phase.  We omit the cool-diffuse and warm-dense phases, which are essentially fixed across all models.}
\end{figure}

The left panel of \fig{massmet_budget} shows that the mass of the CGM is essentially identical across all runs.  The total mass of the CGM is about a factor of 2 times the expected baryon fraction for the halo.  The reason the value is above unity comes from the fact that we are taking all gas within $3 R_{200}$, whereas the baryon fraction is for baryons within the virial radius.  In a run with no feedback (not shown), the CGM mass is a factor of 3 smaller, due to the fact that the gas falls almost monotonically onto the central galaxy.  The total mass of the CGM is insensitive to the changes in the wind strength or heating, suggesting that it reflects some form of equilibrium between accretion and mass ejection.

Though the CGM mass does not vary much, small variations in the mass of the CGM show up as large variations in the mass of the ISM and stars, since the CGM is much more massive.  The mass of the galactic component is clearly more sensitive to feedback, as can be inferred from \fig{SFR}.  The right panel of \fig{massmet_budget} shows the metal content of the galaxy and the CGM.  For all runs, the stars and ISM are significantly metal-enriched above the CGM (which is about 3-10\% solar, with more massive halos having more enriched CGM gas).  The CGM is much more massive than the ISM and stars, so despite a lower metallicity, the total CGM metal mass is comparable to the metal mass in the galaxy.

The metal content of both the CGM and the galaxy is sensitive to the feedback model, and follows the general trend shown in \fig{SFR}.  The runs which have more total star formation up to $z=2$ have more metals in the CGM and galaxy at $z=2$.  The main exception to this trend is the pristine/metal-enriched winds, since these runs explore the parameter space of wind metal-loading.  The Pristine (Fully Enriched) Winds run retains more (less) metals in the galaxy, and correspondingly has less (more) metals in the CGM than the Fiducial run.  

Neither the total gas or metal mass of the CGM is significantly affected by the thermal energy content of the winds.  Our implementation of AGN also does not strongly alter the global mass/metal content of the CGM.  However, as we show in the next section, the stellar and AGN feedback models do affect the physical state of the CGM and the distribution of the metals within it.

\subsection{Density and Temperature of the CGM}
\label{sec:dens_temp}

Motivated by typical conventions in studying the IGM \citep[e.g., ][]{2012ApJ...759...23S}, as well as noting typical CGM densities and temperatures, we define four CGM phases\footnote[2]{We have checked that none of our results depend sensitively on the specific phase cuts used.}:
\begin{itemize}
\item ``cool-diffuse'': $T < 10^5$ K, $n_\text{H} < 10^{-3}$ cm$^{-3}$
\item ``cool-dense'': $T < 10^5$ K, $n_\text{H} > 10^{-3}$ cm$^{-3}$
\item ``warm-diffuse'': $T > 10^5$ K, $n_\text{H} < 10^{-3}$ cm$^{-3}$
\item ``warm-dense'': $T > 10^5$ K, $n_\text{H} > 10^{-3}$ cm$^{-3}$
\end{itemize} 
Here $n_\text{H}$ refers to the number density of H atoms.  Recall that CGM gas is defined to exclude starforming gas, which by construction imposes an upper density limit on the cold-dense and hot-dense phases of $\rho < \rho_\text{SF} = 0.13$ cm$^{-3}$.  

The diffuse phases strongly dominate the volume of the CGM, but about half of the total mass is in the dense gas (the large majority of which has $T <10^5$ K).  Figs. \ref{fig:mass} and \ref{fig:met} show the mass and metal fractions, respectively, of the CGM gas phases.  

The results demonstrate that while feedback has an effect on the mass budgets in the CGM phases, it has a substantially larger impact on the corresponding metal budgets.  

The effect of our AGN implementation, which in the radio-mode inflates large hot ``bubbles", is to heat the CGM, reducing the mass fraction of cool-dense gas and boosting the amount of gas in the warm-diffuse phase.  The ``No AGN'' model, which has the same cold, purely kinetic winds as the Fiducial model, has 60-85\% of all metals in the CGM in the cool-dense phase.

The role of wind-heating in the CGM mass and metal budgets is complex.  Comparing the Fiducial to the Fixed-E Hot Winds model, which has the same specific wind energy but features hotter, slower winds, we find that the Fixed-E Hot Winds have a higher amount of mass and metals in the warm-diffuse CGM (again, at the cost of the cool-dense phase).  This would seem to suggest that at fixed specific energy, the partition between thermal and kinetic energy does impact the CGM phases.  

However, the Faster Winds model and Fixed-v Hot Winds are similar to the above case, in that they have the same specific wind energy, but the Fixed-v Hot Winds have 50\% of this energy in thermal energy.  The only difference is that these models have a higher specific wind energy than the Fiducial model.  The Faster Winds model and Fixed-v Hot Winds have remarkably similar CGM mass- and metal-budgets, despite the fact that the Fixed-v Hot Winds are significantly hotter, and $\sim30\%$ slower than the Faster Winds model!  This correspondence between the two models indicates that wind thermalization is occurring; as the winds come into contact with the CGM, they shock and lose the memory of how the energy was partitioned when it was first launched from the galaxy.  We have checked that the chosen wind particle recoupling criterion affects wind thermalisation significantly less than the parameters of the feedback model (see \sect{converge} for more discussion about the effect of varying the recoupling density).

The difference between the lower-energy pair of runs (Fiducial and Fixed-E Hot Winds) and the higher-energy pair of runs (Faster Winds and Fixed-v Hot Winds) indicates that the process of wind thermalization is more effective at higher specific wind energies.  At lower specific wind energies, thermalization is less efficient, and the CGM retains some memory of how much thermal energy vs kinetic energy was in the wind.  Conversely, at higher specific wind energies, thermalization removes this memory, which leads the Faster Winds (a purely kinetic wind) and the Fixed-v Hot Winds (a run with an equal split between kinetic and thermal energy) to have almost the same CGM mass- and metal-budgets.

The wind mass-loading has no direct effect on the distribution of mass and metals among the various CGM phases.  However, as seen in Sections \ref{sec:sf} and \ref{sec:mass_met}, the wind mass-loading does affect the total metal content in the CGM, by influencing how many stars are able to form.

The amount of cool-dense vs warm-diffuse gas depends weakly on the metal content of the winds.  This is driven by metal line cooling in the halo.  Increased wind metal loading fills the halo with more metals, which increases metal line cooling effectiveness.  This, in turn, helps more gas condense out of the warm-diffuse phase into the cool-dense phase.  This effect can also be seen in terms of slightly increased covering fractions of LLSs and DLAs for higher wind metal loading factors (see Figs. \ref{fig:rudie_172} and \ref{fig:rudie_203}).

\subsection{Radial Gas Profiles}
We now compare the radial structure of halo gas in the different feedback runs.  Note that this includes both ISM at small radii and CGM at larger radii.  See \fig{radprof} for radial temperature and metal profiles.  In both panels of this figure, we take all central galaxies with halo mass $10^{11.8} M_\odot < M_\text{halo} < 10^{12.2} M_\odot$ (there are roughly 50 such halos at $z=2$ in each simulation). 

\subsubsection{Temperature Profiles}
\begin{figure*}
\includegraphics[width=\newFigurewidth]{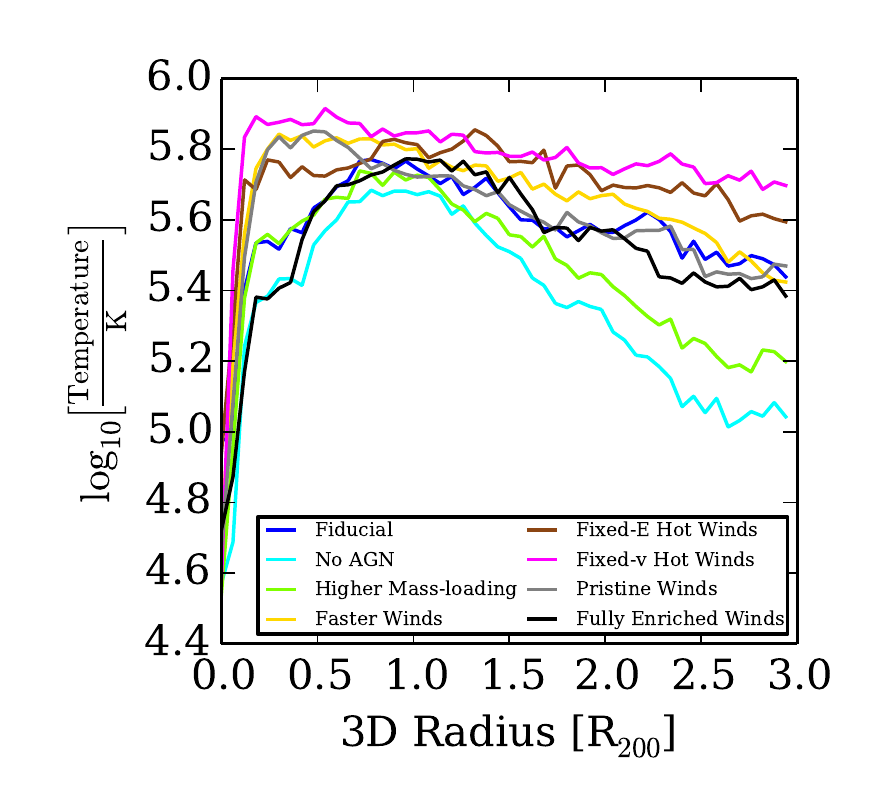}
\includegraphics[width=\newFigurewidth]{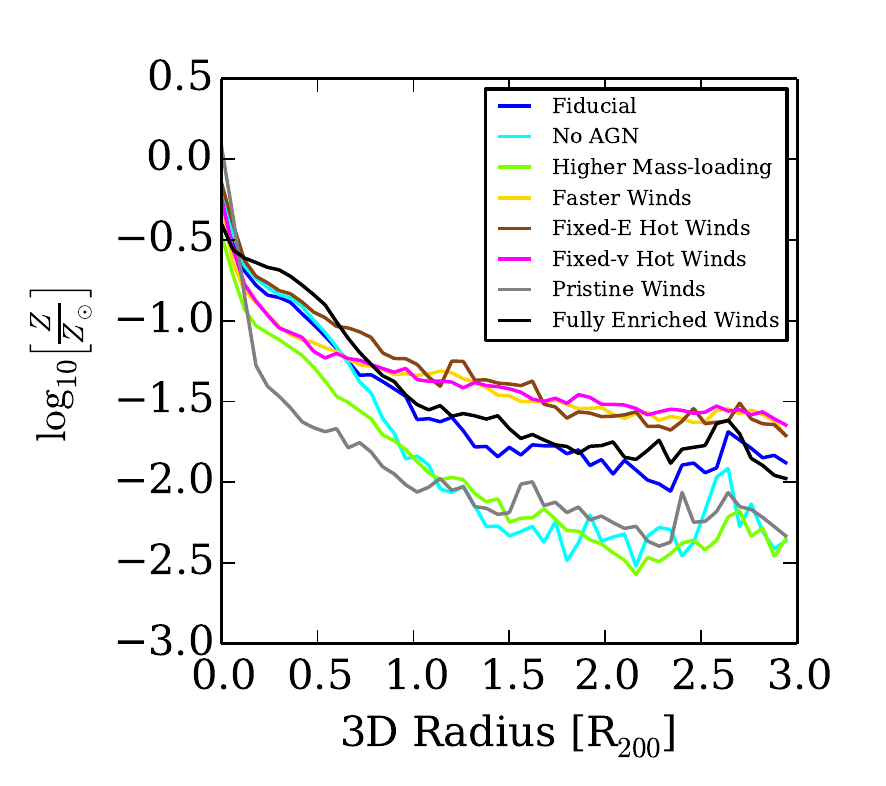}
\label{fig:radprof}  
\caption{Radial mass-weighted temperature (left panel) and metallicity (right panel) profiles around $M\sim 10^{12} \Msun$ halos at $z=2$.  The radius is given in units of the $R_{200}$.  Left panel: Comparing the Fiducial and No AGN runs, we note that AGN feedback significantly enriches halo outskirts.  Wind specific energy also plays a major role; the winds with higher specific energy have flatter and more extended metal profiles, due to a higher IGM enrichment floor from enrichment at early times. Right panel: Wind thermalization occurs at between $0.15 R_{200} \lesssim r \lesssim 0.6 R_{200}$.  Note that higher specific energy winds thermalize and contribute to a higher CGM temperature.  The virial shock is seen at the $r \sim R_{200}$ in all runs, as a convergence in the temperature.  Outside of the virial shock, the AGN dominates the CGM heating, as apparent from the large difference between the Fiducial and No AGN runs.}
\end{figure*}

The left panel of \fig{radprof} shows the mass-averaged gas temperature as a function of radius, in units of the virial radius (which, for these halos is about $98 \pm 8$ proper kpc).  The temperature profiles diverge most strongly inside and outside the virial shock regime.  The CGM gas near the virial shock has an essentially fixed temperature across all simulations.  This temperature is determined by the shocking of infalling material, which by mass dominates over the outflowing winds.

The first regime, at small radii ($0.15 R_{200} \lesssim r \lesssim 0.6 R_{200}$), is the region where winds first reach the CGM.  Wind thermalization first occurs here, as the winds shock upon reaching the ambient halo gas.  This is clear from the Faster Winds model, which has a higher CGM temperature in this regime despite the fact that the wind temperature is identical to the Fiducial run.  In addition the Faster Winds model temperature is similar to the Fixed-v Hot Winds model, which is suggestive since the latter model has hotter winds, but both runs have the same specific energy.  Hence the correspondence in CGM temperature between these two runs with different wind temperatures implies that the intrinsic thermal/kinetic energy partition of the wind is less influential than the total specific wind energy on the CGM temperature inside the virial shock.

Outside the virial shock ($r \gtrsim 1.5 R_{200}$), the AGN dominates the CGM heating, as apparent from the large difference between the Fiducial and No AGN runs.  (Note that the temperature differences among the runs with AGN at $r \gtrsim 1.5 R_{200}$ are comparable to the differences at $r \sim R_{200}$).  The importance of AGN heating at large radii is also seen indirectly in the low temperatures of the High Mass-loading run, since in this run the AGN grow much more slowly (see, for example, Fig. 14 of V13).  This inhibited growth, in turn, hampers the effectiveness of the radio-mode feedback since in our model the radio-mode feedback depends on both the black hole accretion rate and the current black hole mass.  

\subsubsection{Metallicity Profiles}
The Pristine Winds model, which has winds launched with zero metallicity (all the metals are retained by the star-forming gas itself), shows that the metal-enrichment from winds is important at essentially all CGM scales.  Among all runs with enriched winds, however, the CGM metal enrichment is comparable out to about the virial radius.  Beyond this radius, the runs diverge, indicating differences in how far metals are able to propagate out of the halo.  

There are three significant factors for determining the metal content of the CGM for $r > R_{200}$.  The first, perhaps most obvious, factor is the total star formation rate.  For example, the Fixed-E Hot Winds have a higher metallicity than the Fiducial run at virtually every radius.  Similarly, the High Mass-loading run has a significantly lower metallicity than the Fiducial run at virtually every radius.  This can be tied to the fact that by $z=2$ the Fixed-E Hot Winds model has produced about 40\% more stellar mass than the Fiducial run (see \fig{SFR}), and hence significantly more metals.  Conversely, the High Mass-loading run has produced 25\% less stellar mass than the Fiducial run by $z=2$, with correspondingly less metals.  It is unsurprising, then, that, all things being equal, the run with more stars formed will generally have more metals at every radius.

The second major factor is the AGN feedback.  Without AGN feedback, by $z=2$, 30\% more stars form overall in the simulation.  Despite this greatly enhanced star formation, the CGM metallicity at $r > R_{200}$ in the No AGN run is about 0.5 dex below the Fiducial run.  Our AGN model is successfully launching metals beyond the virial radius.  It is interesting to note that, without AGN, the metal content beyond the virial radius is comparable to the Pristine Winds run.  This qualitatively shows how important AGN are for transporting metals outside of the virial radius, in the context of our models.

The third major factor contributing to metals at large radii is the specific wind energy.  The two runs with a higher specific wind energy than the Fiducial run both have lower overall star formation (the Faster Winds and Fixed-v Hot Winds runs have 50\% and 43\% less total stellar mass formed, respectively, by $z=2$).  However, despite the significantly lower total metal production, these two runs both have about 0.4 dex higher metallicity at $r > R_{200}$ than the Fiducial model.  The Faster Winds case is to be expected; faster winds are able to transport metals out of the halo more effectively.  However, comparing the Fixed-v Hot Winds and the Fiducial run, both of which have the same wind speed, shows that the additional thermal energy component helps drive metals out as well (the winds are typically ejected with a temperature hotter than the virial temperature of the halo; see the Appendix for a discussion of how hot the heated winds are).  Furthermore, the Faster Winds and Fixed-v Hot Winds curves lie essentially on top of each other, underlining that at high wind energies, the exact way that the wind energy is partitioned between thermal and kinetic energy is not of great importance.

\subsection{Enrichment as a function of overdensity and redshift}
\label{sec:IGM}

\begin{figure}
\includegraphics[width=\newFigurewidth]{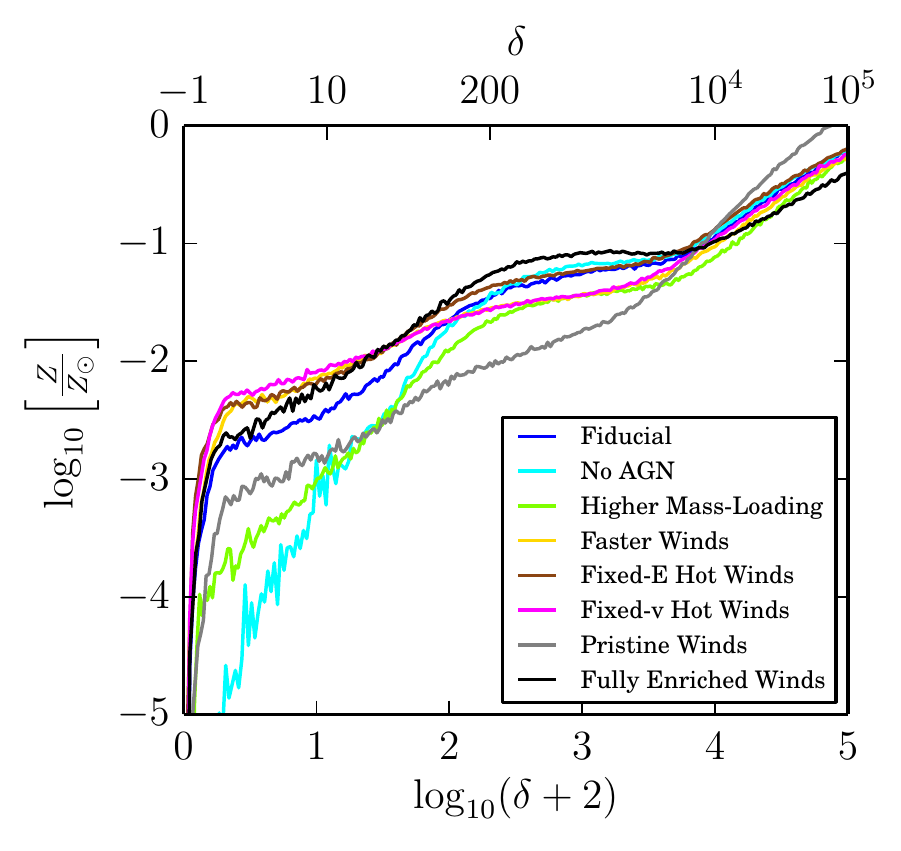}
\label{fig:enrich_od}
\caption{The enrichment as a function of dark matter overdensity at $z=2$.  The curves show the mean mass-weighted metallicity in each bin of overdensity $\delta$.  (Note that since $\delta$ is defined as $\delta \equiv \frac{\rho-\bar{\rho}}{\bar{\rho}}$, the minimum value of $\delta$ is -1, where $\rho=0$.  To plot $\delta$ on a logarithmic plot, we add 2 to $\delta$ so that $\delta=-1$ corresponds to a log$_{10}$ of 0.).  Deep within the halo ($\delta \gg 200$) there is relatively little variation, compared to the CGM outskirts ($\delta \sim 200$) and the IGM ($\delta \ll 200$).  AGN strongly enrich the CGM outskirts and the IGM.  Similarly, high specific-energy winds are able to escape the halo and enrich the IGM.  See \fig{IGM_evol} for a picture of how the IGM enrichment evolves with redshift.}
\end{figure}

\fig{enrich_od} shows the enrichment in the entire simulated box as a function of dark-matter overdensity $\delta$ at $z=2$.  This is calculated by finding the closest dark-matter particle to each gas cell, and computing the dark matter overdensity within a nearest neighbor search of 64 particles.  Studying the metallicity as a function of $\delta$ gives a broader context to the overall metal enrichment in the box, since it includes the IGM as well as the CGM and ISM.  The main limitation of this approach is that the dark-matter overdensities do not inherently contain information about the relevant dark matter halo mass (for example, two dark matter particles with $\delta > 200$ both almost certainly exist inside collapsed halos, however they could belong to halos of very different masses).

\fig{enrich_od} shows the same trends as before, with the added context of the IGM enrichment.  For very high overdensities ($\delta \gg 200$), which correspond to gas deep within the halo, most of the runs have essentially the same level of enrichment.  The primary exceptions are the Pristine Winds and Fully Enriched Winds, which retain or drain the most amount of the ISM metals, respectively.  

The Pristine Winds run dips significantly below the other curves in the range $10^2 \lesssim \delta \lesssim 10^3$, indicating that the outskirts of halos are significantly enriched by the metal-loading of winds (the same trend is seen around massive halos in \fig{radprof}).  However, all of the other runs have roughly comparable metallicities within this regime.  

The main difference between the curves in the right panel of \fig{radprof} is seen in the enrichment outside halos ($\delta < 200$).  Here, AGN clearly dominate the enrichment.  Furthermore, both the higher-energy models (Fixed-v Hot Winds and Faster Winds) have a higher metallicity here, as well as the Fixed-E Hot Winds.

\begin{figure}
\includegraphics[width=\newFigurewidth]{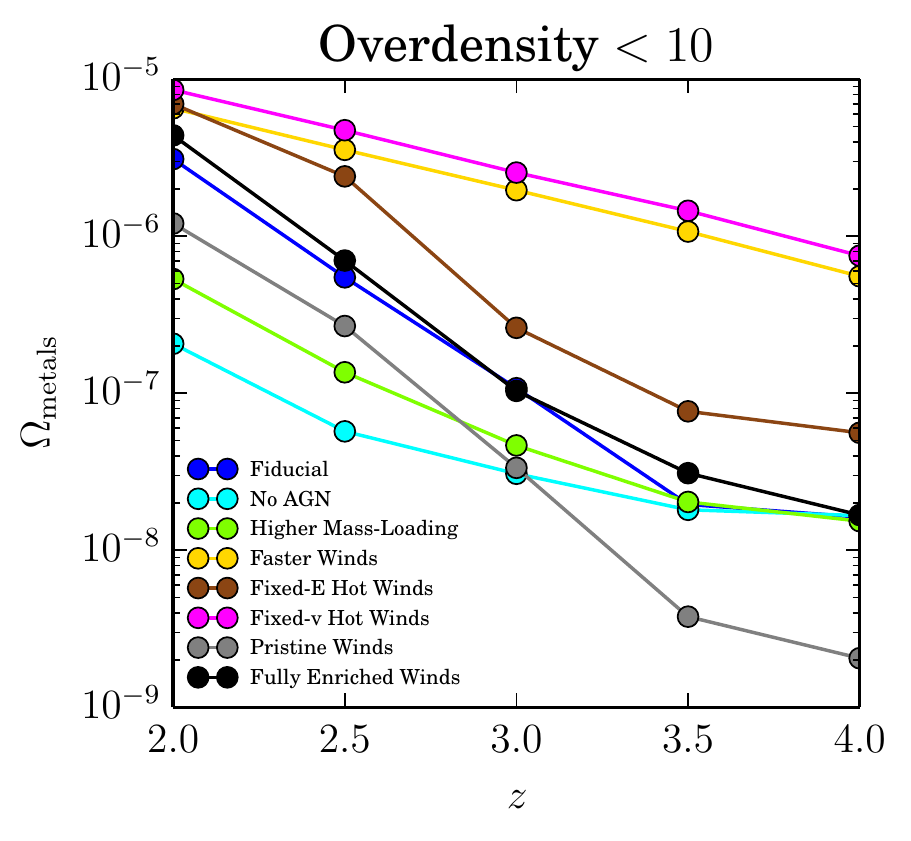}
\label{fig:IGM_evol}
\caption{The enrichment history of the IGM.  Each curve shows the sum of all metal mass in gas with $\delta < 10$, as a function of redshift.  AGN have a major effect in enriching the IGM at relatively late times (after about $z \lesssim 3$).  Conversely, high-energy winds can escape small halos and enrich the IGM at earlier times ($z > 4$), regardless of how the energy is partitioned between thermal/kinetic components.}
\end{figure}

\begin{figure*}
\includegraphics[width=\newFigurewidth]{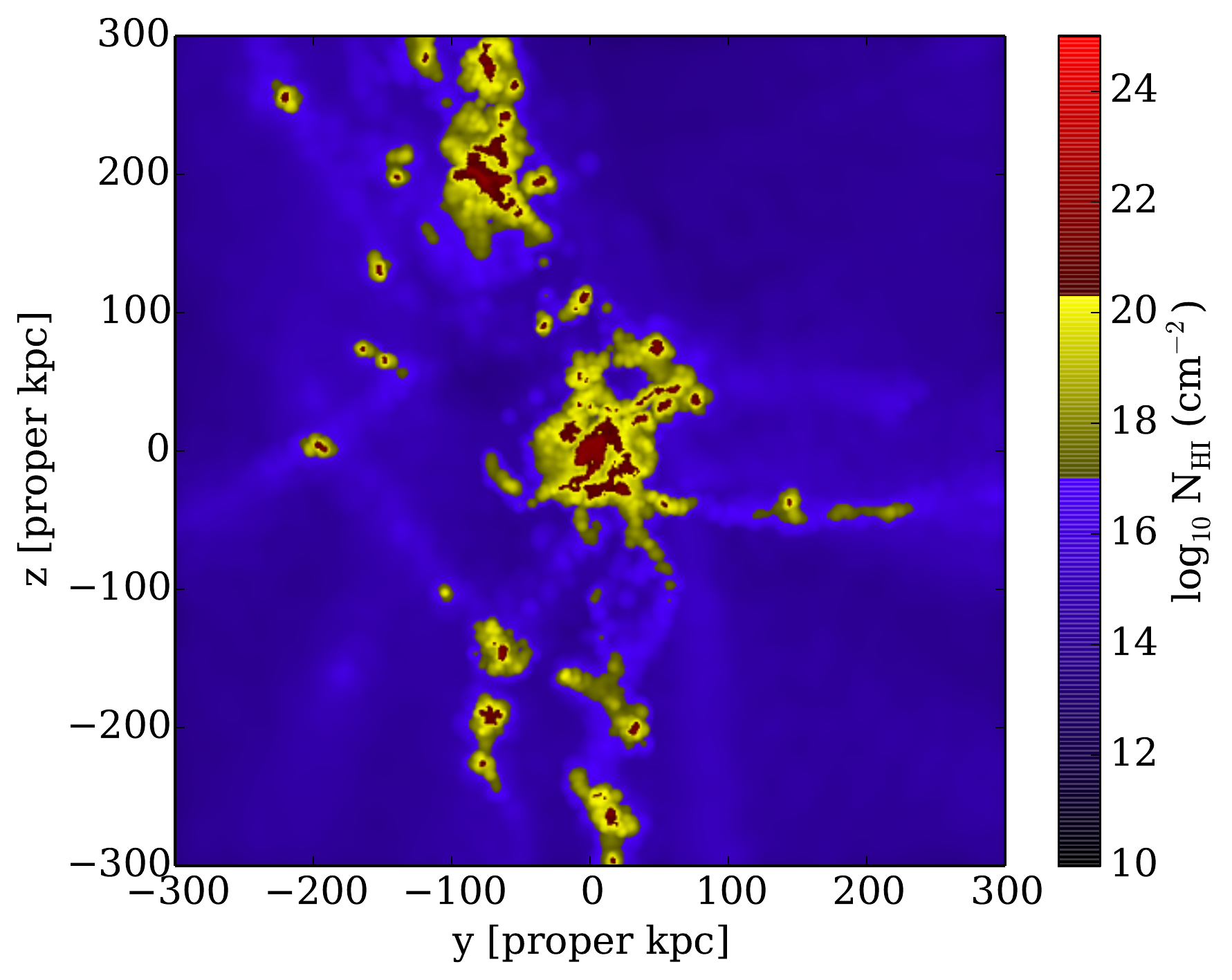}
\includegraphics[width=\newFigurewidth]{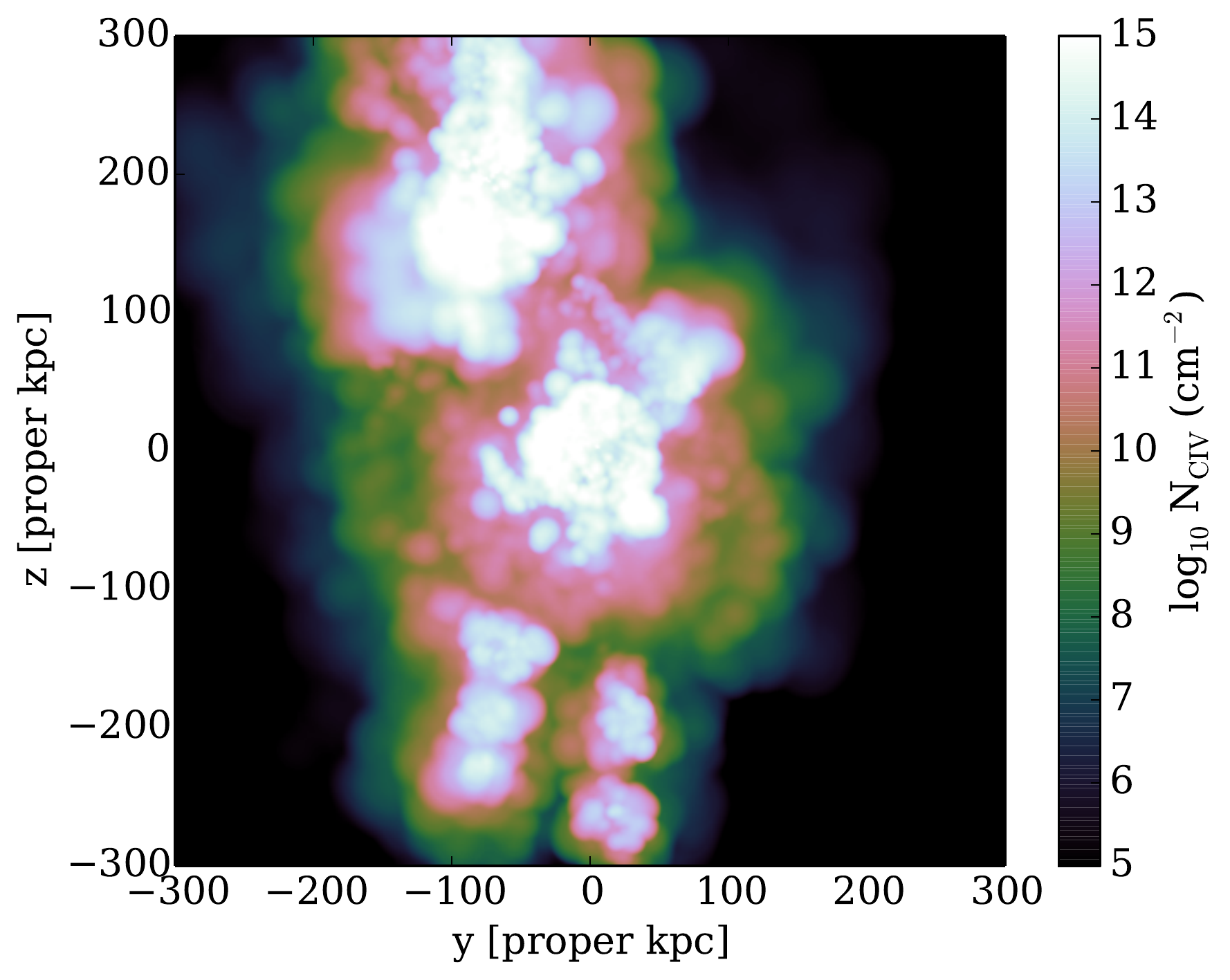}
\label{fig:grids}
\caption{Projected H\ion{I} (left) and C\ion{IV} (right) maps of a single $\sim 10^{12} \Msun$ halo in the fiducial model at $z=2.5$.  The color scheme for the H\ion{I} is selected to highlight LLSs in yellow and DLAs in red.}
\end{figure*}

\fig{IGM_evol} shows the metal content of the IGM versus redshift.  The IGM here is selected as all gas near low-overdensity dark matter ($\delta < 10$).  The most important feature of this figure is that different feedback parameters affect not only the total enrichment of the CGM/IGM gas, but also when that enrichment occurs.  Inspecting the enrichment as a function of redshift clarifies some of the trends seen in Figs. \ref{fig:radprof} and \ref{fig:enrich_od}.

First, we return to the effect of AGN, which again clearly enable many more metals to escape the halo at $z \lesssim 3$ (comparing the Fiducial and No-AGN curves).  As seen before, even the pristine-winds run has a higher IGM metallicity than the run with no AGN (and metal-loaded winds) after $z=3$, showing again the powerful effect that our AGN model has on the enrichment of gas outside halos.  

Secondly, \fig{IGM_evol} highlights the significant role of the wind specific energy.  Both runs with high specific wind energy (Faster Winds and Fixed-v Hot Winds) have a much higher enrichment at $z \gtrsim 2$ than the various low-specific-energy wind models.  Furthermore the profiles are quite flat, indicating that the CGM/IGM metals outside halos in high-specific-energy winds are much older than in low-specific-energy wind models.  

A general picture emerges from these two trends: the high specific energy winds can more easily escape from halos, leading to early enrichment of far-CGM/IGM gas by small halos as they build up their stellar content at $z > 4$.  On the other hand, low specific energy winds cannot easily escape the halo, so in this case the far-CGM and IGM gas is only enriched when AGN become massive enough to begin ejecting metals out of the halo, at $z \lesssim 3$.  This story of cosmic enrichment has observational consequences; specifically we will show in \sect{obs} that C\ion{IV} (and other tracers of the warm CGM such as O\ion{VI}) is an observational probe of these trends.

\section{Observational Constraints and Predictions}
\label{sec:obs}

\subsection{Circumgalactic H\ion{I} at $z \sim 2$}
In \sect{results} we showed how our choice of feedback model parameters affected metal enrichment.  We now compare the model predictions to an important high-redshift observational constraint: circumgalactic hydrogen.  In the following section, we will also show that highly ionized oxygen can probe the differences between the models.

\begin{figure}
\includegraphics[width=\newFigurewidth]{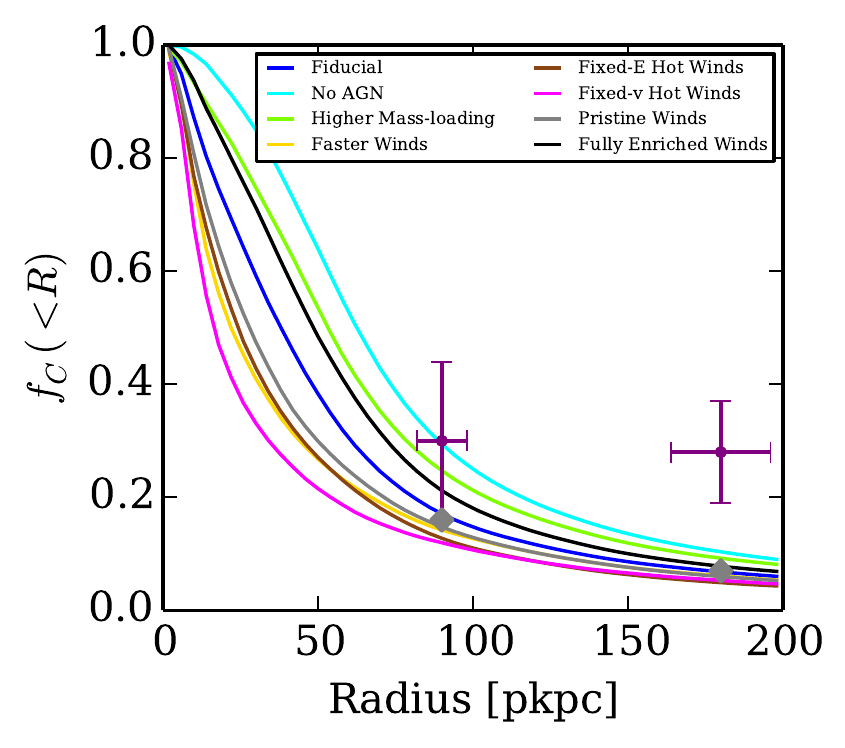}
\label{fig:rudie_172}
\caption{Cumulative covering fractions for $N_\text{HI}>10^{17.2}$ cm$^{-2}$ versus radius in proper kpc (pkpc), around $M \sim 10^{12} M_\odot$ halos, compared to observations from \protect\cite{2012ApJ...750...67R}.  The gray diamonds show the corresponding theoretical values from \protect\cite{2011MNRAS.418.1796F}}
\end{figure}

\begin{figure}
\includegraphics[width=\newFigurewidth]{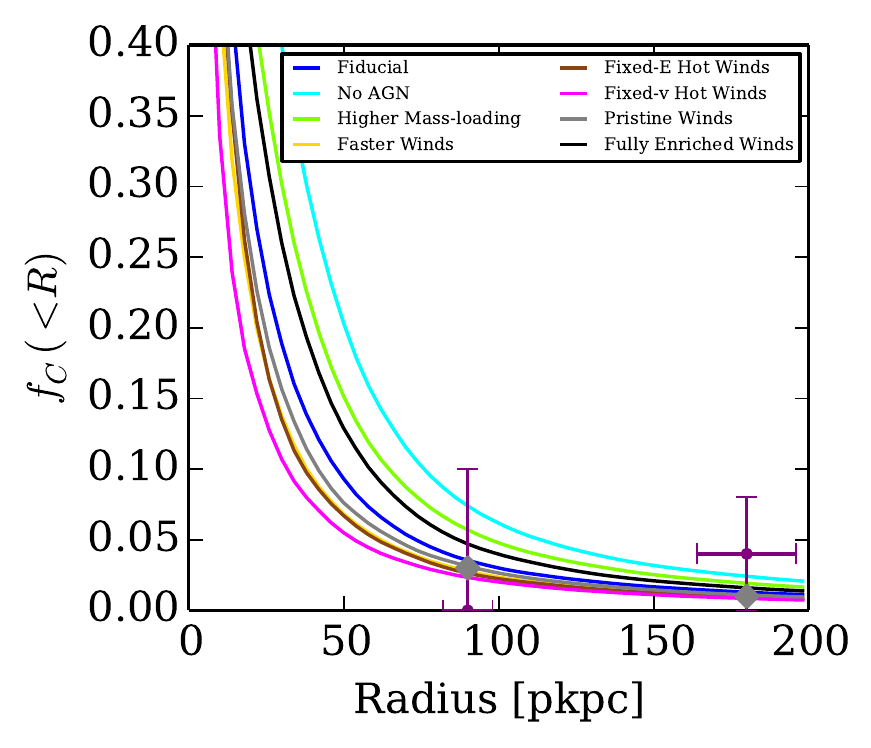}
\label{fig:rudie_203}
\caption{Cumulative covering fractions for $N_\text{HI}>10^{20.3}$ cm$^{-2}$ versus radius in proper kpc (pkpc), around $M \sim 10^{12} M_\odot$ halos, compared to observations from \protect\cite{2012ApJ...750...67R}.  The gray diamonds show the corresponding theoretical values from \protect\cite{2011MNRAS.418.1796F}.}
\end{figure}

In Figs. \ref{fig:rudie_172} and \ref{fig:rudie_203}, we compare the optically thick H\ion{I} covering fractions around massive halos in our simulations to observational data from \cite{2012ApJ...750...67R}, who measured this quantity around star-forming galaxies at $z \sim $ 2--2.5.  \fig{rudie_172} shows the cumulative covering fraction (by cumulative we mean the total covering fraction within a given radius) for sightlines with $N_\text{HI} > 10^{17.2} $ cm$^{-2}$, corresponding to LLS and above.  \fig{rudie_203} shows the same as \fig{rudie_172} but for a column density threshold $N_\text{HI} > 10^{20.3} $ cm$^{-2}$, corresponding to DLAs.    We select central halos with mass $10^{11.8}\Msun < M_\text{halo} < 10^{12.2} \Msun$.  All of our simulations show a modest redshift dependence for the LLS covering fraction (with shallower, more extended profiles at earlier times) while the DLA covering fraction does not evolve significantly over $z \sim 2-2.5$.  

Comparing to a no-feedback run (not shown), we find that feedback is necessary to boost the LLS covering fraction into the observed range \cite[see also][]{FaucherGiguere:2014wu}.  Almost every simulation with purely cold winds is consistent with the LLS covering fraction for $r<R_{200}$.  The only exception is the Pristine Winds run, which has a significantly lower LLS covering fraction, due to insufficient metal line cooling in the less enriched halo gas.  Conversely both runs with heated winds and the run with fast winds have too low a LLS covering fraction at $R_{200}$.  The LLSs thus provide an interesting constraint on how hot the wind can be when it exits the galaxy, as well as on the total specific wind energy.  However, the LLS observational error bars are still too large to definitely rule out any models.

One major caveat, is that no runs produce a large enough cumulative covering fraction for $r \sim 2 R_{200}$.  \fig{rudie_172} shows that all the models converge at large radii, also agreeing with the theoretical results from \cite{2011MNRAS.418.1796F}, and all are about 2 $\sigma$ away from the observational point.  This agreement, despite quite different stellar and AGN feedback models, suggests that it may be difficult to reconcile observations and theory simply by changing the parameters of the feedback model.

There is also the possibility that systematic errors may be affecting the observations, leading to an overestimation of the LLS covering fraction at $r \sim 2 R_{200}$. An effect we do not include is the ability of the bright quasar to drown out the light from nearby small galaxies. By making the true host of the LLS unobservable, this effect could bias the sample towards associating H\ion{I} absorption with larger galaxies further away instead of smaller galaxies close by, thus artificially boosting the measured absorption at large distances from large galaxies. \cite{2014MNRAS.438..476P} found that line blending may also artificially boost the LLS covering fraction, an effect which would likely be strongest near the outskirts of halos, where the LLS covering fraction is dropping rapidly.

Contrary to the $N_\text{HI}>10^{17.2}$ results discussed above, \fig{rudie_203} shows that the observational error bars on DLAs are too large to distinguish between the models.  Perhaps most interesting here is that AGN do have a significant effect on DLAs around massive halos, though they do not have much effect on the overall DLA population since most DLAs inhabit smaller halos where AGN feedback is less efficient \citep{2014arXiv1405.3994B}.  See also \cite{2014arXiv1405.3994B}, who study the DLA properties of our feedback model in much more detail, and finds that it achieves good agreement with observational constraints on the abundances and properties of DLAs.

\subsection{Circumgalactic C\ion{IV} at high redshift}
\label{sec:CIV_z3}

\begin{figure*}
\includegraphics[width=\newFigurewidth]{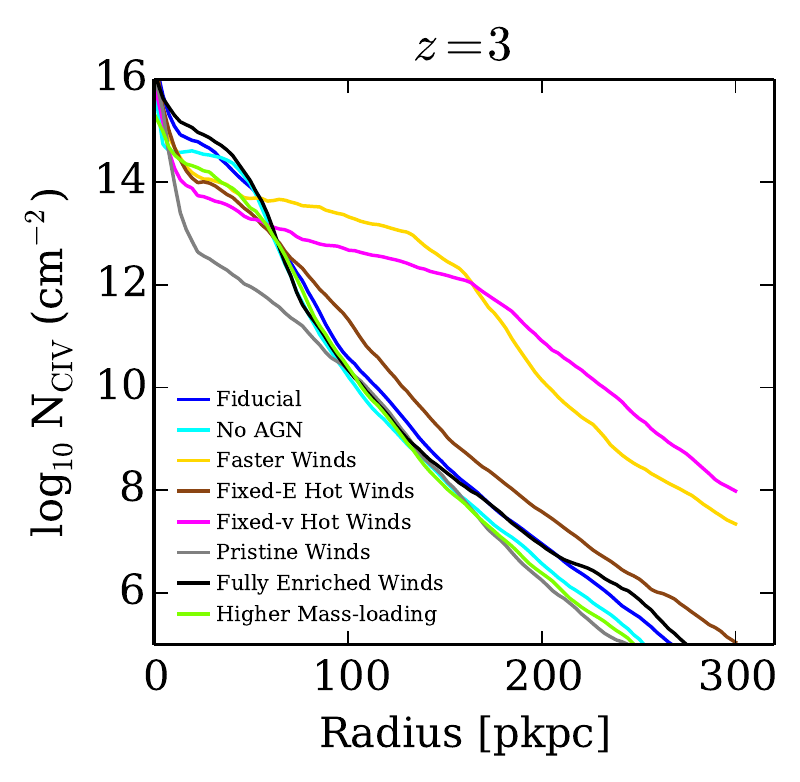}
\includegraphics[width=\newFigurewidth]{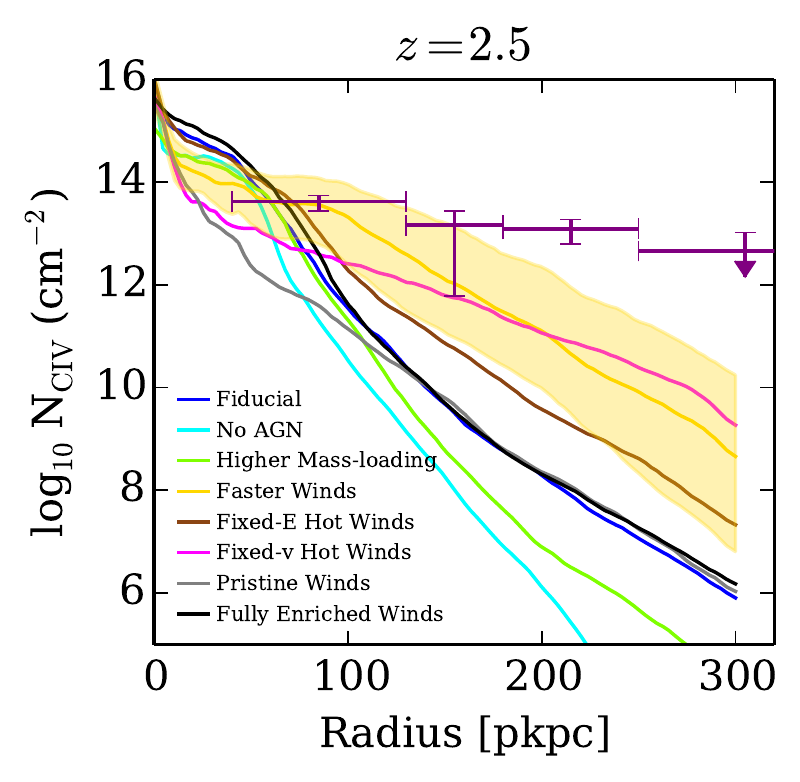}

\label{fig:CIV}
\caption{C\ion{IV} column density in $M \sim 10^{12} M_\odot$ halos at (Left) $z=3$ and (Right) $z=2.5$ versus radius in proper kpc (pkpc), compared to observations from \protect\cite{2014MNRAS.445..794T}.  The yellow shaded region shows the 32nd and 68th percentile ($1-\sigma$) of the Faster Winds model, to show the typical spread about each of these curves.  At $z=3$, AGN have not yet significantly heated the CGM. The effect of the wind specific energy is however substantial, suggesting that circumgalactic C\ion{IV} may be a sensitive tracer of specific energy in galactic winds. At $z=2.5$, AGN feedback increases the amount of C\ion{IV} $100-300$ kpc from the galaxy, complicating interpretation.
However, all simulations have significantly less C\ion{IV} at these radii than is seen observationally.  The models that come closest to the observations are those with sufficiently high specific wind energy to launch metals beyond the virial radius.  However, these models also produce overall too few stars at late times (see \fig{SFR}).}
\end{figure*}

The analysis of cosmic enrichment showed that wind specific energy and AGN both have an important role in determining when the CGM outskirts ($R > R_{200}$) and IGM are enriched.  High specific energy winds enrich at early times from small halos, whereas low-specific-energy wind models enrich at later times, due to the effect of AGN growth.  By $z\sim 2$, the two effects can become confused.

However, at earlier times ($z \gtrsim 3$), our results make the clear prediction that there should only be extended CGM metal profiles in models with high-specific-energy winds (regardless of wind energy partitioning between thermal and kinetic energy).  We find that this theoretical trend is reflected in the observational tracer of C\ion{IV} and other tracers of the warm CGM, such as O\ion{IV}.  We focus on C\ion{IV} in this section since there are already good observations of this ion at high redshift, whereas the shorter-wavelength O\ion{IV} doublet is contaminated by the Lyman-$\alpha$ forest.  \fig{CIV} shows the median column density of C\ion{IV} in the CGM of $M \sim 10^{12} \Msun$ halos at $z=2.5$ and $z=3$.

The high-redshift ($z \gtrsim 3$) C\ion{IV} column density is a good tracer for galactic stellar-driven winds since AGN have not yet had time to strongly heat the CGM.
\fig{CIV} shows that at $z\gtrsim 3$, there is a clear distinction between the C\ion{IV} profiles for high specific energy winds and low specific energy winds.  Both runs with high specific energy winds have much more extended C\ion{IV} profiles, with about $N_\text{CIV} \sim 10^{12}$ cm$^{-2}$ out to 150 kpc.  In the same radial range, we find $N_\text{OVI} \sim 10^{14}$ for the high energy winds, which may be suggestive, as \cite{2011Sci...334..948T} found that at $z \sim 0$, virtually all star-forming galaxies have a near 100\% covering fraction of $N_\text{OVI} \sim 10^{14}$ cm$^{-2}$ absorbers.

At the slightly lower redshift $z=2.5$, the effects of AGN are clearly visible in \fig{CIV} by comparing the fiducial model with the no AGN curve.  AGN act to boost the amount of C\ion{IV} at all radii.  However, all the models show significantly less extended C\ion{IV} profiles than that observed in \cite{2014MNRAS.445..794T}.  For example, the median C\ion{IV} column density at $\sim 200$ kpc for the fiducial model is roughly five orders of magnitude below the observed point!  Even if we make the unphysical assumption that all of the carbon is found in the C\ion{IV} phase, the fiducial model still underpredicts the observed C\ion{IV} column densities at $\sim 200$ kpc.  

The models with higher specific wind energy (Faster Winds and Fixed-v Hot Winds) are closer to the observations but still severely underpredict the C\ion{IV} profile at $\sim 200$ kpc.  Furthermore, this better agreement comes at a cost, since these higher energy models produce less stars overall in the box, moving these models slightly out of the observed star-formation rate density values as a function of redshift around $z \sim 2$ (see \fig{SFR}), and even more so at lower redshift.  Thus we find a strong tension between the requirements of extended metal profiles at high redshift with the constraints of global star-formation.  Despite the various physics variations tried in this work, none of our models satisfies both observational constraints.  This highlights the fact that CGM observations are an independent and stringent constraint on galactic feedback models.  

One possible resolution to this problem could be to introduce a two-velocity wind model, instead of the single-velocity wind models we have considered in this work.  The lower velocity wind would, as in the Fiducial model, not escape the halo and recycle back into the galaxy at later times.  This recycling would maintain a sufficiently high overall star-formation in the simulation to match the observed star-formation rate density. On the other hand, the higher velocity wind component could escape the halo at early times and generate the extended metal-line profiles seen observationally.  This type of multi-component wind may better represent the properties of multi-phase winds seen in higher resolution simulations such as \cite{2012MNRAS.421.3522H}.

We plan to further address properties of the warm CGM phase, and specifically whether the gas is collisionally ionized or photo-ionized, in forthcoming work.

\section{Discussion and Summary}
\label{sec:discussion}

In this work, we have used the moving-mesh code \textsc{arepo}, with phenomenological feedback implementations tuned to match numerous statistical galactic properties, especially at $z=0$ \citep[see V13 and ][]{2014MNRAS.438.1985T}, to examine the corresponding predictions for the high-redshift circumgalactic medium.  We inspect large-scale cosmological volumes, with a statistical sample of galaxies, using eight feedback variations which compare the relative effects of AGN feedback and wind speed, specific-energy, thermal-kinetic energy partition, and metal-loading.  Our main conclusions from this work are as follows:
\enumerate{
\item The fiducial feedback model of V13 produces good overall agreement with observations of circumgalactic H\ion{I} around $M \sim 10^{12} \Msun$ halos at high redshift.  The observed DLA covering fraction, unfortunately, has error bars which are larger than the variation between the different runs.  The LLS covering fraction, on the other hand, is sensitive to both the wind heating as well as total specific wind energy.  The V13 model matches the LLS covering fraction at $R_{200}$, but all models underpredict the observed covering fraction at $2 R_{200}$.

\item Conversely, the C\ion{IV} column densities around similar objects ($M \sim 10^{12} \Msun$ halos at high redshift) are much more difficult to match.  While most of our simulated models match the observations at $\sim 100$ kpc, the observed metal profiles are much flatter. The observed C\ion{IV} column density at $\sim 200$ kpc is more than an order of magnitude higher than the corresponding median column densities in our models.  The models which come closest to the observations have high specific wind energy, and are thus in tension with observations of the star-formation rate density.

\item AGN feedback plays a major role in the enrichment of circumgalactic and intergalactic environments.  AGN play little to no role before $z \sim 3.5$ but after $z=3$, AGN significantly enrich the CGM and IGM\footnote[2]{Note that in principle, the specific redshift at which the AGN begin to dominate is a statement which is dependent on box-size, since more massive BHs can form earlier in larger boxes.  Nevertheless, we find the same redshift trend in the Illustris simulation, which has a factor of 27 more volume than our box \citep[for example, see the CGM/IGM enrichment history shown in Fig. 26 of][]{2014arXiv1405.2921V}}.  They play such a role that for $z \leq 3$, even a run with zero-metallicity winds and AGN has greater IGM enrichment than the run with fiducially enriched winds but no AGN!  The main reason for the effectiveness of AGN feedback in our model is the radio mode, triggered at low accretion rates, which effectively heats the CGM with large bubbles.  In other words, ignoring AGN or simply replacing AGN feedback with an ad-hoc quenching mechanism may result in a skewed understanding of the enrichment of the CGM. 

We caution, however, that our AGN radio mode feedback model is likely too vigorous at ejecting gas; for example \cite{2014arXiv1405.3749G} finds that the most massive halos at $z < 1$ have lost far too many baryons due to the radio mode feedback.  However, the problem of halos losing too much gas diminishes with smaller halo mass (especially $M < 10^{12.5} M_\odot$) and with higher redshift.  Comparing two runs with and without radio mode feedback, we confirm that the effect of large-scale loss of baryons in the halo is negligible at the halo masses and redshifts of interest in this paper (for example, see left panel of \fig{massmet_budget}).

\item Metal line cooling is an important effect in the CGM, which we have shown by varying the metal-loading of winds and checking explicitly with a run that has only primordial cooling.  Increased wind metal loading increases the metallicity of the halo gas, which in turn leads to more metal line cooling.  This metal line cooling, in turn, condenses more halo gas into the cool-dense phase.  This effect leads to a slight boosting of the circumgalactic LLS and DLA covering fraction.

\item The specific wind energy (wind energy per unit mass) of galactic outflows significantly affects the enrichment of galactic environments.  In this work we have explored two specific and related ways in which the enrichment of the CGM and IGM is altered by the specific wind energy of the galactic feedback:
\begin{itemize}
\item The radial extent to which metals can propagate.
\item The time at which the IGM is enriched.
\end{itemize}

We now summarize both of these effects.

\textit{The radial extent to which metals can propagate.}  Winds with higher specific energy enrich to significantly larger radii than winds with lower specific energy.  More extended metal profiles might be expected in the case of a higher-velocity wind.  However, we find that even a hotter wind launched at the same velocity as a cold wind will deposit its metals further (since the hotter wind has a higher total specific energy, it can more easily escape the potential well of the halo.)

Note that given a fixed budget of total supernova energy, winds with higher specific energy must necessarily have a lower mass-loading.  The mass-loading of winds has little direct effect on how well the winds can escape the halo, and only impacts CGM enrichment indirectly by altering the amount of stars which can form.

\textit{The time at which the IGM is enriched.}  Since winds with higher specific energy can more easily escape the halo, the IGM is enriched much earlier in a high energy wind model than in the low-energy wind models.  Even at $z = 4$ the high-energy-winds IGM is 1-2 orders of magnitude more enriched than the lower-energy-winds IGM, since winds arise from small halos as soon as they begin to form stars.  This effect also causes the metal profiles around halos to be much flatter than those with cold winds, since the background IGM metallicity is much higher.  Once AGN feedback becomes important ($z \lesssim 3$), AGN-driven outflows can also escape the halo, which leads the low-energy-wind models to a similar level of enrichment as the higher-energy-winds model by $z \sim 2$.  Hence, the effects of high-energy galactic winds and AGN can be differentiated by when they enrich the IGM; high-energy winds at $z >4$ and AGN at $z \lesssim 3$. 

\item Wind thermalization is more efficient at high wind energies.  At low wind energies, the way the wind energy is partitioned between kinetic and thermal components does have some effect on, for example, the temperature and density of the CGM metals.  However, at higher wind energies, the energy partition becomes inconsequential; the CGM retains little memory of how the wind energy is partitioned, since the wind shocks and thermalizes efficiently with the ambient halo gas.  Wind thermalization typically occurs within roughly $0.15 R_{200} \lesssim r \lesssim 0.6 R_{200}$.

Given the broad parameter space we have investigated in this work, we can begin to connect the somewhat disparate pictures of the CGM found in other recent theoretical studies.  Specifically, our work sheds light on the differences between the theoretical CGM properties found in \cite{Ford:2013wk,2013MNRAS.432...89F} and \cite{2012ApJ...760...50S,2013ApJ...765...89S}.  The winds in \cite{2012ApJ...760...50S,2013ApJ...765...89S} are clearly escaping at high speed (typically about 250 km/s, but up to 800 km/s for some parts of the wind), and the outflows are also quite hot due to the blastwave feedback prescription.  This implies a high specific energy of the winds.  In this case, our models predict that metals can easily escape the virial radius, leading to an early enrichment of the IGM at $z>4$ by small halos; this is precisely what \cite{2012ApJ...760...50S,2013ApJ...765...89S} find.  

Conversely, \cite{Ford:2013wk,2013MNRAS.432...89F} employs cold winds with an enforced velocity scaling that typically launches winds below the escape speed (similar to the Fiducial model in this work), implying that the winds have a significantly lower specific energy than the \cite{2012ApJ...760...50S,2013ApJ...765...89S} model.  In this case, our models predict that most of the metals remain embedded within the virial radius, likely leading to a strong cycle of re-accretion of recently ejected winds; again, this is what  \cite{Ford:2013wk,2013MNRAS.432...89F} find.  Furthermore, since they employ an artificial quenching mechanism in massive halos instead of an effective AGN feedback model \citep{2013MNRAS.434.2645D}, it is possible that the metal re-accretion and outflow cycle is even further exaggerated, since in our model, AGN are instrumental at ejecting metals out of massive halos.

Since we infer that the \cite{Ford:2013wk,2013MNRAS.432...89F} wind models are lower specific energy than that of the blastwave model in \cite{2012ApJ...760...50S,2013ApJ...765...89S}, our results imply that wind thermalization in the former model is likely less effective, and that the CGM retains more memory of how cold or hot the winds were when ejected.  This may explain why the hot-wind model of \cite{2012ApJ...760...50S} predicts that the lowest over-densities are enriched by relatively cold gas, whereas the cold-wind model of Ford et al. showed exactly the opposite trend; the lower the over-density the higher the ionization state of the gas \citep[][]{Ford:2013wk,2013MNRAS.432...89F}.  The oldest metals in the hot-winds case are metals that were ejected out of the halo in a hot wind at early times, have cooled, and are now associated with the outskirts of the CGM.  Conversely, in the cold-winds case, metals tend to be recycled in the halo much more than in the hot-winds case; therefore new CGM metals tend to be colder (since they were recently in a cold wind) than the old CGM metals which have managed to stay in the halo without being recycled.

The strong effect of both AGN and galactic wind energy on the enrichment of the CGM means, unfortunately, that our lack of knowledge about feedback creates an uncertain theoretical picture of the enrichment history of galactic environments.  The silver lining is that the strong effect of feedback on the enrichment of the CGM and IGM gives us new observational tools with which to constrain galactic feedback, beyond simply using star formation or galactic morphology, which have degeneracies in constraining feedback.  In this work we identified two specific observational predictions.  The first is that the LLS covering fraction around galaxies is likely a probe of galactic feedback.  The second is that the C\ion{IV} profiles around $\sim 10^{12} \Msun$ halos at $z \gtrsim 2.5$ are a sensitive tracer of galactic wind energy, which we find generates an interesting tension between circumgalactic C\ion{IV} at high redshift with the observational constraints on the star-formation rate density.  We anticipate that as observational data of the CGM continues to improve, the CGM will prove to be a powerful constraint on the details of galactic feedback.}

\section*{Acknowledgements}
JS thanks Rob Simcoe, Kate Rubin, Laura Sales, Vicente Rodriguez-Gomez, and Dylan Nelson for their helpful comments.  VS acknowledges support by the European Research Council under ERC-StG EXAGAL-308037.  LH acknowledges support from NASA grant NNX12AC67G and NSF grant AST-1312095.


\bibliographystyle{mn2e}
\bibliography{CGM_bib}

\appendix
\section{Convergence Properties}
\label{sec:converge}
\begin{figure}
\includegraphics[width=\newFigurewidth]{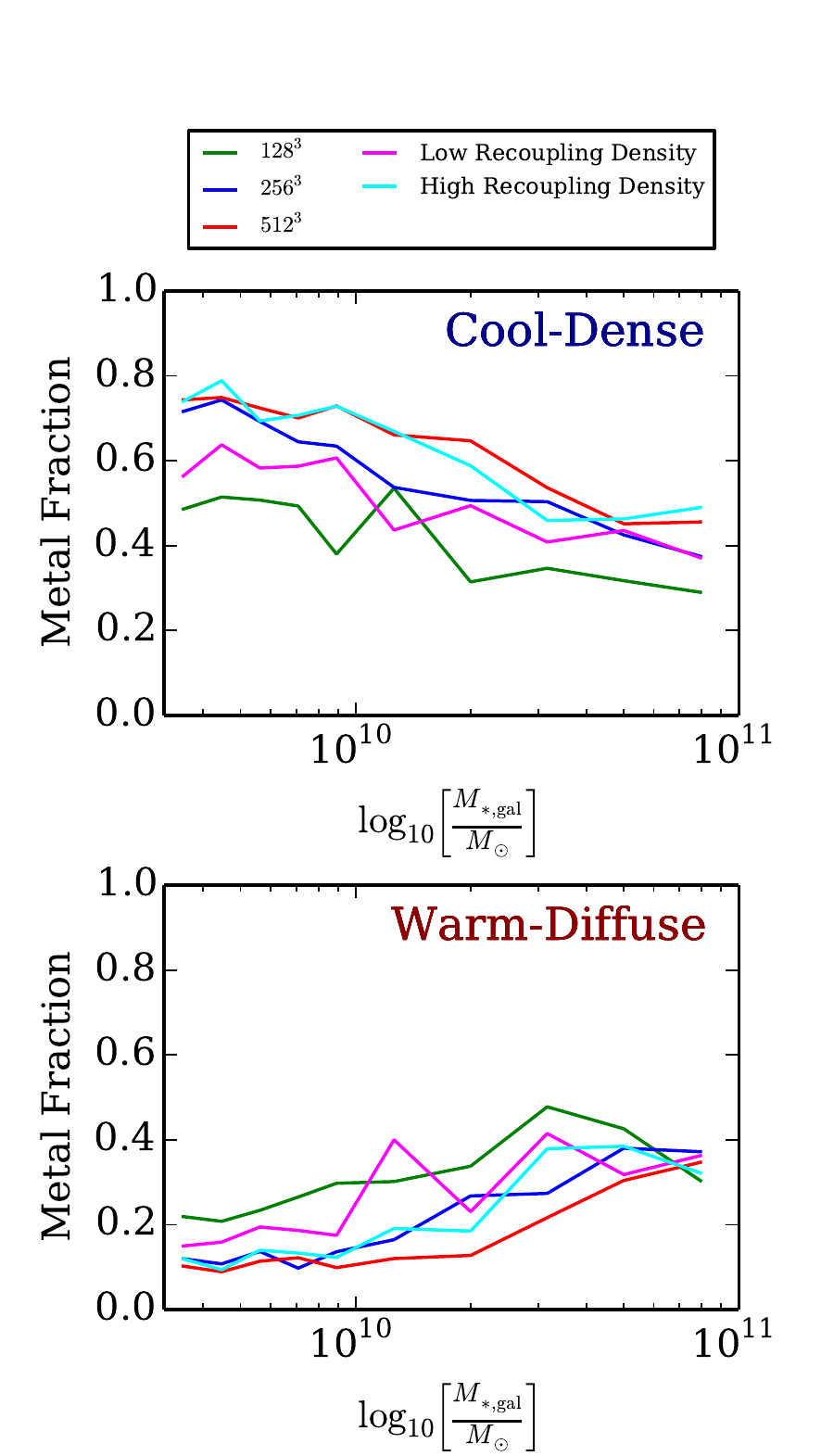}
\label{fig:z_budget_res}
\caption{Convergence of the CGM metal budget with resolution.  Each curve shows a simulation with a different number of resolution elements. The green, blue, and red curves vary the resolution, with the label showing the number of gas resolution elements.  $256^3$ is the resolution used throughout this study.  The cyan and magenta curves show, at the $256^3$ resolution, the effect of varying the wind recoupling density.}
\end{figure}

\fig{z_budget_res} shows the convergence of the CGM metal budget with resolution.  We choose to show this quantity as we found it to be the least well-converged of the CGM 
properties we examine. However, even here the difference between $256^3$ (the resolution used throughout the rest of this study) and $512^3$ is small compared to, say, the effect of AGN feedback (shown in \fig{met}). Convergence for the CGM is as good as that for the star formation rate, discussed in V13 and \cite{2014MNRAS.438.1985T}. This validates our use of these simulations in this work.

\fig{z_budget_res} also shows the effect of our wind recoupling scheme on the CGM metal budget.  The recoupling density is the density at which the hydrodynamically decoupled winds are recoupled to the hydrodynamics scheme.  The cyan and magenta curves in \fig{z_budget_res} show two runs with the fiducial physics model and resolution, but with the wind recoupling density halved and doubled.  We note that a higher recoupling density effectively lowers the energy of the wind, since the wind shocks into the higher density region and radiatively cools more of its energy away.  This is seen in \fig{z_budget_res}, where increasing the recoupling density slightly increases the amount of cool-dense gas and reduces the amount of warm-diffuse gas, a similar effect to that of increasing the energy of the winds, seen in \fig{met}.  Hence, we find that we explore roughly the same parameter space by simply keeping the recoupling density fixed and varying the wind energy, which is the approach we have adopted in this work.  Furthermore, the net effect of changing the recoupling density is negligible compared to the overall spread among the different feedback models explored in this work.

\section{How hot are the winds?}
\label{sec:append}

\subsection{Parametrizing the partition between thermal and kinetic energy}
The total energy produced by a new stellar population, divided by the mass of the population is given by the following equation:
\begin{equation}
E_\text{tot} = \text{KE} + \text{TE} = \frac{\eta}{2} v^2 + \eta u_\text{th}
\end{equation}
where $\eta \equiv \frac{\dot{M}_\text{wind}}{\text{SFR}}$ and $u_\text{therm}$ is the thermal energy per unit mass in the wind.  Given the observational trend that $v_\text{wind}$ tends to scale with the halo dispersion velocity, we can parametrize both $v$ and $u_\text{therm}$ in terms of the halo 1D dispersion velocity $\sigma$:
\begin{eqnarray}
&& v^2 = \alpha_v \sigma^2    \\
&& u_\text{th} = \alpha_\text{th} \sigma^2
\end{eqnarray}

Now, let us define the ratio of energy in the thermal component as $\gamma$, so $\text{TE} = \gamma E_\text{tot}$ and $\text{KE} = (1-\gamma) E_\text{tot}$.  Then we can find $\gamma$ in terms of the above parameters:

\begin{equation}
\frac{\text{KE}}{\text{TE}} = \frac{(1-\gamma) E_\text{tot}}{\gamma E_\text{tot}} = \frac{\frac{\eta}{2} \alpha_v \sigma^2}{\eta \alpha_\text{th} \sigma^2} = \frac{\alpha_v}{2 \alpha_\text{th}}
\end{equation}

Rearranging, we find:
\begin{equation}
\gamma = \frac{2 \alpha_\text{th}}{2 \alpha_\text{th} + \alpha_v}
\end{equation}

\subsection{Comparison to the virial temperature}
Assume winds are energy-driven, with all energy from the supernovae going into winds.  Now, as above, assume that some fraction $\gamma$ goes into the thermal energy of the winds, while the remaining energy goes into the kinetic energy.

Assume that the total supernova energy per unit stellar mass released from a single stellar population is $E_\text{tot}$.  Then, following \cite{2012MNRAS.426..140D}, we find that the increase in temperature $\Delta T$ that occurs when adding the thermal component of available energy is given by:
\begin{equation}
\Delta T = (\gamma_\text{gas}-1) \frac{\mu m_\text{H}}{k_\text{B}} \left[ \gamma E_\text{tot} \right] \frac{m_*}{m_\text{g,heat}},
\end{equation}
where $m_*$ is the initial mass of the particle and $m_\text{g,heat}$ is the total gas mass which the heat is dumped into.  Note that $\frac{m_*}{m_\text{g,heat}}$ corresponds to $\eta^{-1}$, since the mass of heated gas is precisely the mass of wind particles generated by the activity of this newly-formed star particle.  Thus,
\begin{equation}
\Delta T = (\gamma_\text{gas}-1) \frac{\mu m_\text{H}}{k_\text{B}} \left[ \gamma E_\text{tot} \right] \eta^{-1},
\label{eqn:deltat2}
\end{equation}

Recalling relations from the above section, $\text{TE} = \gamma E_\text{tot} = \eta u_\text{th}$.  Rearranging, we can isolate $\eta$:
\begin{equation}
\eta^{-1} = \frac{u_\text{th}}{\gamma E_\text{tot}} = \frac{\alpha_\text{th} \sigma^2}{\gamma E_\text{tot}}
\end{equation}

Plugging in to \eq{deltat2}, we find:
\begin{equation}
\Delta T = (\gamma_\text{gas}-1) \frac{\mu m_\text{H}}{k_\text{B}} \left(\alpha_\text{th} \sigma^2 \right)
\label{eqn:deltat3}
\end{equation}

Recall the definition of the virial temperature of a halo: $T_\text{vir} = \frac{\mu m_\text{H}}{2 k_\text{B}} v_\text{vir}^2$, where $v_\text{vir}$ is the virial velocity, defined as the circular velocity at the virial radius.  Taking the ratio of $\Delta T$ and $T_\text{vir}$, we find:
\begin{equation}
\frac{\Delta T}{T_\text{vir}} = 2 (\gamma_\text{gas}-1)\alpha_\text{th} \left(\frac{\sigma}{v_\text{vir}}\right)^2
\end{equation}

The relationship between $\sigma$ and $v_\text{vir}$ is not simple, since it depends on the density profile of the halo, which is in this case affected by baryonic physics.  We find that over a wide range of halo masses, $\left(\frac{\sigma}{v_\text{vir}}\right)^2 \simeq \frac{2}{3}$.  If we take $\gamma_\text{gas} = \frac{5}{3}$, then we find that $\frac{\Delta T}{T_\text{vir}} > 1$ for $\alpha_\text{th} > \frac{9}{8}$. 

In this work we choose $v = 3.7 \sigma$; this is the same parametrization used in \cite{2013MNRAS.436.3031V}, who showed that stellar properties of galaxies are sensitive to the choice of wind speed.  This scaling choice for $v$ gives $\alpha_v = 3.7^2 = 13.7$.  Given this choice for $\alpha_v$, the criterion $\alpha_\text{th} > \frac{9}{8}$ implies $\gamma > 0.14$.  In other words, for our choice of wind velocity scaling, if an additional energy component, consisting of more than 14\% of the total supernova energy, is put into heating the winds (as is true in the case of Fixed-v Hot Winds, which have $\gamma = 0.5$), the resulting winds temperatures will exceed the virial temperature of the host halo.
\label{lastpage}
\end{document}